\documentclass{aa}  
\usepackage{graphicx}
\usepackage{natbib}
\usepackage{amsmath}
\bibpunct{(}{)}{;}{a}{}{,}
\usepackage{txfonts}
\begin{document}
   \title{Detection efficiency and photometry in supernova surveys}

   \subtitle{-- the Stockholm VIMOS Supernova Survey \rm{\bf{I}} 
   \thanks{Based on observations collected at the European Organisation
   for Astronomical Research in the Southern Hemisphere, Chile, under ESO 
   programme ID's 167.D-0492(C) and 167.D-0492(D)}}

   \author{J. Melinder\inst{1}
          \and
          S. Mattila\inst{2,3}
		  \and
		  G. \"Ostlin\inst{1}
		  \and
		  L. Menc{\'i}a Trinchant\inst{1}
		  \and
		  C. Fransson\inst{1}
          }

   \offprints{J. Melinder}

   \institute{Stockholm Observatory, AlbaNova University Centre,
	      SE-106 91 Stockholm, Sweden\\
              \email{jens@astro.su.se}
         \and
             Tuorla Observatory, University of Turku, V\"ais\"al\"antie 20,
			 FI-21500 Piikki\"o, Finland\\
		 \and
		 	 Astrophysics Research Centre, School of Mathematics and
			 Physics, Queen's University Belfast, Belfast BT7 1NN, UK
             }

   \date{Received ; accepted }

  \abstract
   {}      
   {The aim of the work presented in this paper is to test and optimise
	supernova detection methods based on the \emph{optimal image
	subtraction} technique.  The main focus is on applying the
	detection methods to wide field supernova imaging surveys and
	in particular to the Stockholm VIMOS Supernova Survey (SVISS).}
  {We have constructed a supernova
   detection pipeline for imaging surveys. The core of the pipeline
   is image subtraction using the ISIS 2.2 package. Using real data
   from the SVISS we simulate supernovae in the images, both inside
   and outside galaxies. The detection pipeline is then run on the
   simulated frames and the effects of image quality and subtraction
   parameters on the detection efficiency and photometric accuracy
   are studied.}
  {The pipeline allows efficient
   detection of faint supernovae in the deep imaging data. It also
   allows controlling and correcting for possible systematic effects in
   the SN detection and photometry. We find such a systematic effect
   in the form of a small systematic flux offset remaining at the
   positions of galaxies in the subtracted frames. This offset will
   not only affect the photometric accuracy of the survey, but also
   the detection efficiencies.}
  {Our
   study has shown that ISIS 2.2 works well for the SVISS data. We
   have found that the detection efficiency and photometric accuracy
   of the survey are affected by the stamp selection for the image
   subtraction and by host galaxy brightness. With our tools the
   subtraction results can be further optimised, any systematic effects
   can be controlled and photometric errors estimated, which is very
   important for the SVISS, as well as for future SN searches based
   on large imaging surveys such as Pan-STARRS and LSST.}


   \maketitle
%

\section{Introduction}
\label{sec:intro}
Imaging surveys to find and characterise variable sources are used
to study a number of different astronomical phenomena. Detection and
photometry of variable sources can be done by subtracting images
taken at different epochs from each other and analysing the subtracted
frames, a technique known as \emph{difference imaging}. In addition
to supernovae (SNe), this technique has been used by various authors
investigating many different subjects, among which are variable stars
in crowded fields \citep[e.g.][]{2006AJ....132.1014C}, gravitational
micro-lensing \citep[e.g.][]{2001MNRAS.320..341A,2006ApJ...636..240S} 
and planet detection \citep[e.g.][]{2007ApJ...655.1103H}.

The most complicated step in the difference imaging process is
taking into account the spatial and temporal
variability of the point spread functions (PSFs) of the images.
The difference imaging technique used in this work is based on 
the Optimal Image Subtraction (OIS) code first presented in \citet[
hereafter AL98]{1998ApJ...503..325A}, later improved upon in \citet[
hereafter A00]{2000A&AS..144..363A} and available in the ISIS 2.2
package. In these papers the code is tested
on variable stars (i.e. point sources with varying brightness with
little or no background light). When detecting SNe the situation can
be quite different.  In this type of data the background light from
the host galaxy often dominates the total flux at the location of
the supernova.

Difference imaging, and OIS in particular, has been used to detect
SNe in many projects. Recently a number of large SN surveys, mainly
aiming at finding thermonuclear SNe for determining cosmological
parameters, have released their first results. Image subtraction
using the OIS algorithm followed by detection of SN candidates
in the subtracted frames is a technique that has been used by
the ESSENCE survey \citep[the observations and data analysis are
described in][]{2007astro.ph..1043M}, the Supernova Legacy Survey
\citep{2006A&A...447...31A} and \citet{2007MNRAS.382.1169P}.
\citet{2005A&A...430...83C} and \citet{2008A&A...479...49B}
used the same technique in their SN surveys aiming to determine
the core collapse SN rate at intermediate redshift. Difference
imaging has also successfully been used in near-infrared SN searches
\citep[e.g.][]{2001MNRAS.324..325M,2002A&A...389...84M} concentrating
on individual starburst galaxies. For surveys with the aim of finding
supernova rates the detection efficiency is the most important issue,
while the SN photometry is of somewhat less importance. However, for
surveys targeting Ia SNe the photometry is of utmost importance since
the photometric accuracy directly determines how well the cosmological
parameters can be constrained. Therefore, in these programmes the
photometry is treated very carefully and not necessarily obtained
from the subtracted frames. For studies of individual SNe difference
imaging has been used to obtain accurate photometry of supernovae with
a complicated background.  This is particularly important at late
times when the SN has become too faint compared to the background
host galaxy in the unsubtracted image for PSF photometry
to give reliable results \citep[e.g.][]{2002A&A...386..944S}.
More recently, difference imaging has also successfully been
applied to mid-infrared SN observations with the Spitzer Space
Telescope \citep{2006ApJ...649..332M} and to adaptive
optics assisted detection of SNe in luminous infrared galaxies
\citep{2007ApJ...659L...9M}.

In this paper we investigate several aspects of difference imaging with
ISIS 2.2 when used to detect and study supernovae. Previous papers have
already reported some tests done on the image subtraction \citep[e.g.,
AL98; A00;][]{2007AN....328...16I} and on the SN detection process
\citep[e.g.][]{2001MNRAS.324..325M}.  The surveys for thermonuclear
SNe all have extensive sections where the methods to ascertain the
photometric accuracy (both in terms of statistical and systematic
errors) is described \citep[e.g.][]{2006A&A...447...31A}. However, no
systematic tests on the subtraction method applied to supernova data
have been reported. The main purpose of this paper is to describe
our supernova detection pipeline and to study how it performs
on wide field imaging surveys, in particular the Stockholm VIMOS
Supernova Survey (SVISS). The SVISS is an imaging survey aiming to
find core-collapse supernovae at $0.1< z < 1.0$ \citep[for details
see][]{1999A&A...350..349D}. The main scientific goal for this
project is to find reliable core collapse supernova rate densities
for this redshift range. We thus need to have detailed knowledge of
the supernova detection efficiency. In this project the photometric
accuracy is important to correctly type the SNe based on colour and
light curve information. Another goal of this paper is to investigate
how supernova detection and photometry (with ISIS used as the image
subtraction tool) are affected by the selection of subtraction and
detection parameters and the presence of a host galaxy.  In this paper
we discuss the subtraction and detection method, in the following
papers we will present the light curves of the detected SNe, the
typing methodology and calculations of SN rates at high redshifts.

The first part of the paper (section \ref{sec:method}) contains a
detailed description of the standard image subtraction method used in
this work. In section \ref{sec:data} we describe the data sets, in
section \ref{sec:sim} how the simulations of supernovae have been
setup and in section \ref{sec:results} we present the results from
the extensive testing.  Finally, we discuss the results and possible
implications in section \ref{sec:disc} and conclude by giving a short
summary in section \ref{sec:summary}. The Vega magnitude 
system has been used throughout the paper.

\section{Image subtraction and SN detection method}
\label{sec:method}
We have constructed a supernova detection pipeline suitable for
large images and multi-epoch data. The pipeline consists of a number
of IRAF/PYRAF scripts that are run in sequence, and go through the
following steps: (i) accurate image alignment over the
entire frame; (ii) convolving the better seeing image to the same PSF
size and shape as the poorer seeing image, using a spatially varying
kernel; (iii) subtracting the images; (iv) detection of sources in the
subtracted frames, using both source detection software and eye-ball
detection; (v) photometry and construction of light curves of the
detected objects.

The input to the pipeline consists of one reference image and a list of
later epoch images.  The reference image is subtracted from each of the
later epoch images and objects with varying brightness are detected
in the first search epoch image. The output from the pipeline is a
list of supernova candidates detected in the first search epoch and
a light curve constructed from measurements on the detected position
in the subsequent epoch images. The detection part of the pipeline
can then be rerun with the previous second epoch as detection epoch
(enabling SNe that have appeared between the first two epochs to be
found). This can be repeated for all search epochs, i.e. once a
particular search epoch has been run through the pipeline, that epoch
can then be used as a reference epoch for subsequent epochs. This is
useful when a search epoch image is of better quality (in terms of
seeing and depth) than the reference epoch image.

\subsection{Registering images}
\label{sec:regmeth}
The subtraction method is quite sensitive to how well the images
are registered. Spatial variability of the convolution kernel
can somewhat compensate for a non-perfect image alignment \citep[also
discussed in][]{2007AN....328...16I}, but in general the images should
be aligned with high precision before running the actual subtraction.
Accurate image alignment is particularly important when the
seeing differences are small, since narrow convolution kernels
will not be able to compensate for poor image alignment.

All the input images are aligned to the reference image by using the
IRAF tasks \texttt{geomap} and \texttt{geotran}. A number of bright
sources visible in all of the frames are used as input coordinate lists
for these tasks.  The exact positions of the sources are found by using
the IRAF task \texttt{imcentroid}. The number of reference sources
used is dependent on the order of the geometrical transformation and
the size of the image.

\subsection{PSF matching and image subtraction}
\label{sec:sub}
The atmospheric conditions will affect the width and shape of the
PSFs of the images. The combination of the atmospheric effects
with the optics, focusing and camera setup results in variations
of the PSF of the images both in spatial coordinates and over time
(i.e. different epochs can have different PSFs, and also different
spatial variation). A successful image subtraction technique must
be able to match the PSF of the reference image to the PSF of the
later epoch images also allowing the convolution kernel to vary
spatially. The OIS technique presented in A00 allows this.

Following A00 (using the same notation) we summaries the OIS
technique. The principal difficulty with image subtraction is to find a
convolution kernel ($K$) that can transform the PSF of a reference
image ($R$) to the PSF of an arbitrary image ($I$). The best-fit
kernel can be found by minimising the sum over all pixels
\begin{equation} 
\label{eqn:kernelsum} 
\sum_i([R \otimes K](x_i,y_i) - I(x_i,y_i) + bg(x_i,y_i))^2,
\end{equation}
where $bg(x_i,y_i)$ is a spatially varying background term.
The kernel can be written as a sum of basis functions
\begin{equation}
K(u,v) = \sum_n a_n(x,y) K_n(u,v)
\end{equation}
where $a_n$ contain the spatial variations of the kernel, which are
polynomial functions of a given degree. $K_n$ are
the basis functions for the kernel and $u$ and $v$ denote the PSF kernel
coordinates. The basis functions are Gaussian functions of the type
\begin{equation}
K_n(u,v) = e^{-(u^2+v^2)/2\sigma_k^2} u^i v^j
\end{equation}
with the generalised index $n=(i,j,k)$. The basis functions can
have different widths (the $k$ index) and different orders (the $i$
and $j$ indices). To deal with a spatially varying background a
polynomial background term is present in Eq. ~\ref{eqn:kernelsum}
and is also included in the solution which is found by solving the
resulting linear system (for details see A98 and A00). 
The reference image is convolved with the kernel and subtracted
from the search image. 

The default parameters for image subtraction used for the SVISS images
are listed in Table~\ref{table:params}.  The basis functions described
above used in ISIS 2.2 are three Gaussian functions of orders 6, 4 and
3. Both \citet{2007AN....328...16I} and \citet{2002A&A...381.1095G}
find that the use of 3 Gaussian functions of these orders works well,
we use these basis functions in our subtractions. We have, however,
done some tests with different widths of the functions which is
reported in section~\ref{sec:widthresults}. The order of the polynomial
functions used for the spatial variation of the kernel in A00 is
2. \citet{2002A&A...381.1095G} use a zero order spatial variability,
but compute local kernels for subregions in the image. For the SVISS
data set we subdivide the large 2k $\times$ 2.5k pixels VIMOS frames
into 9 subregions and use a first order spatial variation of the
kernel. We have chosen to use kernel and stamp sizes scaled for the 
poorer seeing image (i.e. the search image), however
we have not found these sizes to affect the subtraction quality
substantially. A very small kernel size (smaller than
the FWHM of the image) will likely cause the subtraction to fail.
A too large kernel size will on the other hand include more noise in
the fit and probably make the subtraction quality worse. In general,
varying the parameters listed in Table ~\ref{table:params}, have
considerably smaller effect on the resulting subtraction quality
than the selection of stamps, data quality and host galaxy brightness
(see section~\ref{sec:tsetup}).

In ISIS a number of sources present in both reference and search image
are used as the data that goes into the kernel solving procedure. A
number of high S/N-sources are selected from the image and a
small subimage (so called stamp) is obtained for each of them. The
selected sources do not have to be point sources, but they have to be
non-saturated and well below the non-linear response regime of the
detector. The stamps are selected by dividing the image into a number
of rectangular sections (as many as the wanted number of stamps). In
each of these sections the brightest non-saturated pixel is found
and the stamp is centred on that pixel. To check that no variable
objects are selected, the fluxes inside the stamps in the two images are
compared and the variable stamps are rejected. 

Our initial test runs showed some problems with the automatic stamp
selection.  The stamp selection is very sensitive to how saturated/bad
pixels are handled in the data, if not treated correctly saturated
objects could be selected as stamps. When simulating bright SNe we
find that the simulated sources themselves are often selected as stamps
even though they are only present in one of the frames, and cause the
subtraction quality to deteriorate. This indicates that the rejection
routine described above is not working for our data set with
simulated SNe. However, often the number of SNe per image is expected
to be quite low, thus the effect on real survey data is expected to
be marginal. In our simulations, where a large number of artificial
SNe were placed in the images the effect is clearly noticeable.

To gain a better control over the stamp selection process
we have modified ISIS to accept an external stamp list. This
version of ISIS has previously been used for SN detection in
\citet{2001MNRAS.324..325M,2004NewAR..48..595M,2007ApJ...659L...9M}
with manual stamp selection. In this study the stamps are selected by
running SExtractor 2.4.4 \citep{1996A&AS..117..393B}
on the reference image, which allows us to reject stamps that are too
faint and to place constraints on the stellarity of the sources. In
this way we can also reject stamps that are suspected to contain a
variable source and then rerun the subtractions. The stamp region list
can also be compared with the bad pixel masks for both the reference
and the detection image to make sure that no stamp regions contain
saturated pixels.

In supernova surveys the sources selected as stamps are not necessarily
point sources because the number of available bright point sources is simply too
small. The majority of the automatically selected stamps will be extended
sources, so if point sources are preferred the external stamp list method
has to be used. In section \ref{sec:sttest} we have examined the effects of
different selection criteria for the stamps and how the number of
stamps affects the subtractions.

\begin{table} 
\caption{Default subtraction parameters}             
\label{table:params}      
\centering                          
\begin{tabular}{l c}        
\hline\hline                
Parameter & SVISS data set value \\    
\hline 
Subregions & 9 \\
Total number of stamps ($n_{st}$)& $\sim 450$ \\
Size of the kernel ($\times$FWHM) & 4.0 \\

Full size of the stamp ($\times$FWHM) & 6.0 \\
Fitting order for the background & 2 \\
Degree of kernel spatial variation & 1 \\
Orders of the Gaussian basis functions & 6/4/3 \\
\hline                                   
\end{tabular} 
\end{table}
\begin{figure*}
\centering
\includegraphics[width=0.75\textwidth]{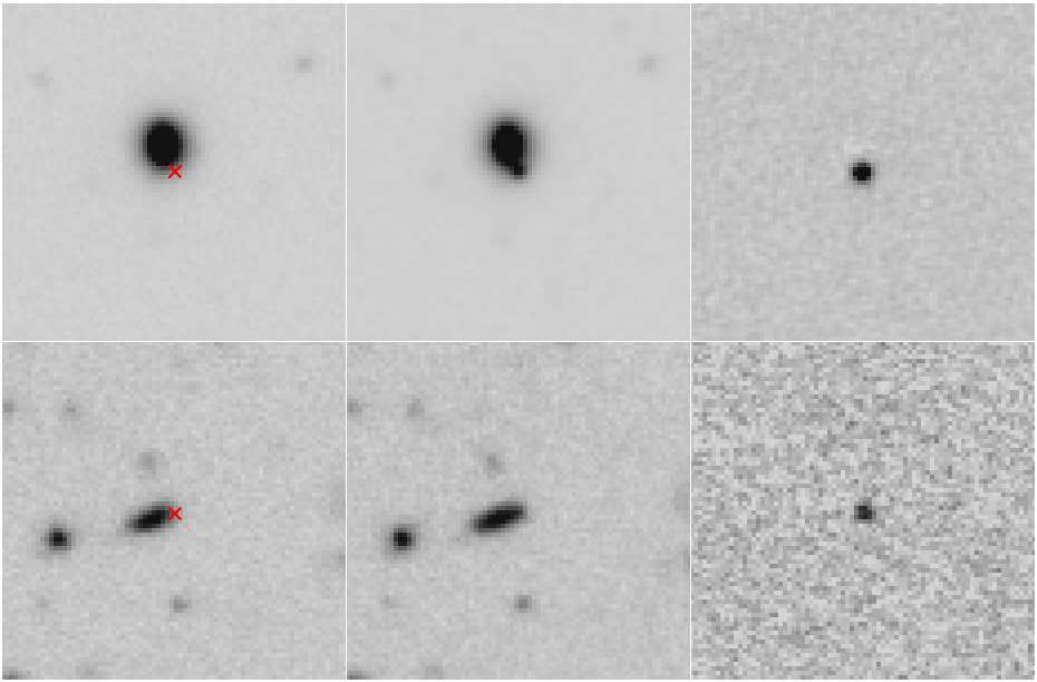}
\caption{Two example supernova candidates detected in the SVISS.
Upper and lower panels show the two candidates with the leftmost panels 
showing the host galaxy in the reference image, the middle panels
the host galaxy in the detection epoch and the rightmost panels the
supernova candidate in the subtracted frame. The crosses in the leftmost
panels mark the positions of the candidates. The candidate in the upper
panel has an $I$ band magnitude of 22.7 and in the lower
panel 24.8. See text for more details.}
\label{figure:snsamp} 
\end{figure*}
\subsection{Source detection}
\label{sec:subdet}
Source detection is done in the subtracted frames using SExtractor (SE) to
get an initial source list. The frames are also manually inspected
to check for possible image defects and obvious spurious detections
(e.g., sources close to saturated objects). The detection thresholds
in SE are set by giving the minimum area of detected objects in pixels
(MIN\_AREA) and the minimum flux for a pixel to be considered detected
(DET\_THRESH). We have used MIN\_AREA parameter corresponding to
a circular area with diameter of roughly $1 \times FWHM$. The DET\_THRES
parameter is a multiplier to the measured pixel to pixel noise (rms) in
the frame, this parameter has been adjusted to detect faint sources
while still keeping the number of spurious detections down (see
section \ref{sec:spur} for a discussion on this). Convolving 
the images affects the noise and therefore also the optimal detection
threshold to use, and since we are using other methods to ensure that
spurious detections are excluded we have chosen to use a low detection
threshold ($1.5 \times$ the rms measured in the subtracted frame).

In SE weight-maps in the form of variance maps can be used to correct
the detections for varying noise levels over the image (due to
e.g. vignetting and exposure time differences). This is especially
helpful in the edges of images. Using the SE option of WEIGHT\_TYPE
set to MAP\_WEIGHT our exposure time maps are scaled to inverted
variance maps within SE. The inclusion of a weight-map in the detection
decreases the amount of spurious detections close to the edges of the
subtracted images considerably.  For each epoch weight-maps in the
form of exposure time maps can be constructed (see \ref{sec:data}
for a description on how this was done for the SVISS data set). For
counts-per-second CCD images where the instrumental noise is negligible
(i.e. the noise is dominated by photon noise from the source or sky)
the standard deviation of the individual pixel flux scales as the
square root of the exposure time. The pixel noise in the subtracted
image is approximately given by quadrature addition of the reference
and search image noise (AL98). Convolving the reference image slightly
changes the noise characteristics of the subtracted image (discussed
in A00), but for constructing the combined weight-maps we have found
it sufficient to use the unconvolved reference exposure time maps. The
combined weight-maps are then computed as follows:
\begin{equation}
W_c = \frac{1}{\frac{1}{W_{ref}} + \frac{1}{W_{s}}},
\end{equation}
where $W_{ref}$ and $W_{s}$ are the exposure time maps (weight-maps)
for the reference image and the search image respectively and $W_c$
is the weight-map for the subtracted image. Note that the weight
maps are only used in the detection step in SE, photometry is done on
the unweighted frame.

All of the detections with valid photometry (in this case
errors lower than 1.0 mag) are kept. The coordinates of the detected
sources are all put into a master candidate list, this is then used
for the photometry (described in the next section). The master
candidate list can also be used to do multi-epoch and multi-band
detection of the sources found in the search epoch image under
consideration. The master candidate list will contain a large
number of spurious detections (due to the low detection threshold
used and the presence of subtraction residuals). These objects can be
excluded by combining three different criteria; (i) too faint,
the sources are fainter than a predefined limit (e.g. 
signal to noise limit from simulations), (ii) 
only present in one filter, (iii) light curve not consistent with 
SN templates.

\subsection{Photometry of variable sources}
\label{sec:phot}
Photometry on the detected sources was done using the IRAF
\texttt{daophot} package. PSF photometry was done on all the detected
candidates using the task \texttt{allstar}. The PSF model used for each
subtracted frame was
constructed from the original worse seeing image (the search
image) using the task \texttt{psf}. In this task point
sources are fitted with model profiles and
the profile yielding the smallest residuals is chosen as the
best PSF fit. The order of variability of the PSF model was set to zero,
i.e. the PSF was not allowed to vary over the frame. Note that the model
was constructed from the real data, thus the shape of the PSF is not
symmetrical and does contain noise. As discussed earlier, often the PSF vary
over the frames. However, using a higher order variability in the
psf photometry did not result in any gain in the photometric accuracy.
We also used aperture photometry with aperture corrections
computed from the original worse seeing image, using the IRAF task
\texttt{phot}. This does give fairly
good results, but seems to be more susceptible to residual flux from the
background galaxies than the PSF photometry, thus giving larger errors
(both statistical and systematic) in general.

Using the pixel to pixel noise in the subtracted image
will not give a correct estimate of the photometric errors, the
pixels will be positively correlated due to the convolution of
the reference image. Another problem in using the pixel to pixel
noise estimate is that it will in general be computed in an annulus
outside the aperture (or outside the size of the PSF model in the
case of PSF photometry). The annulus region for most SNe will be
outside the host galaxy, at least in the SVISS case where most
host galaxies are just barely resolved. The noise estimate in this
region can in principle be quite different from the noise at the
SN position where residuals from the subtracted host galaxy will
also be present. Both these effects result in the noise estimates
given by \texttt{daophot} and \texttt{phot} underestimating the true
photometric errors severely. Instead we use the simulations described
in section~\ref{sec:sim} to obtain reliable error estimates. By
simulating SNe at different brightness in each epoch and finding the
scatter in their measured magnitudes, we obtain an estimate of the
true photometric error for each point on the SN light curve.

The limiting magnitude for a given epoch depends on the depths of both
the reference and the search images. Since the noise in the subtracted
image will be approximately equal to quadrature added noise of the
images, the subtracted image will in general have a higher level of
noise by a factor $\sqrt{2}$ and therefore be $\sim 0.4$ magnitudes
less deep than the individual images. For the SVISS data, this
is true for the pixel-to-pixel noise. The subtracted frames do have
on the order of $\sqrt{2}$ higher pixel rms (but slightly lower due to
the convolution). However, we have found that the limiting magnitudes
computed for the individual SVISS epochs based on pixel-to-pixel noise
($m_{sim,rms}$ in Table~\ref{table:data}) are not representative of the
true depth as measured by by our simulation technique. We have 
measured the photometric scatter of the simulated SNe in the
unsubtracted epoch images and obtained more conservative depth estimates
($m_{sim,sim}$ in Table~\ref{table:data}). 

The reason why the photometric scatter estimate results in
larger errors than the simple pixel-to-pixel estimate is a combination
of several factors. We believe that the main factor is that our deep
images are coming close to the confusion limit. In photometry the
subtraction of a local background becomes very complicated when the
background is allowed to vary over sizes comparable to the aperture
size (or fitting radius for the PSF magnitudes). The background sources
just at the confusion limit inside the photometric aperture/fitting
region and background annulus will vary from one simulated source to
the another and can thus cause the scatter from the measured magnitudes
to go up. Incorrect background subtraction in the presence of large
scale background variations also seem to be a contributing factor,
although of lesser importance than the one described above.  It is
worth noting that subtracting the images actually serves to remove
these two factors making the photometric scatter go down. In reality
this effect is countered by the increase in pixel-to-pixel noise caused
by the subtraction, see the discussion in section~\ref{sec:pos}.


\section{The SVISS data sample} 
\label{sec:data} 

In this section we describe the data set that was used to investigate
the efficiency of the subtraction method. The data were obtained
with the VIMOS instrument \citep{2003SPIE.4841.1670L} mounted on
the ESO Very Large Telescope (UT3) at several epochs. The reference
epoch was obtained in autumn 2003 and the search epochs autumn
2004. The VIMOS instrument has four CCDs, each 2k$\times$2.4k
pixels with a pixel scale of 0.205\arcsec/pxl, covering a total
area of $4\times$56 sq. arcmin. The SVISS observations were
obtained in two fields, covering parts of the Chandra Deep
Field-South \citep[][]{2001ApJ...551..624G} and the ELAIS-S1
field \citep{2004AJ....127.3075L}, and in 5 broad band filters
($U$, $B$, $V$, $R$ and $I$). The supernova search filters are $R$
and $I$, with roughly twice the exposure time in $I$ compared to
$R$. Observations in these filters have been divided into 1 reference
epoch (henceforth epoch 0) and 7 search epochs.  The time difference
between epoch 0 and epoch 1 is about one year and the search epochs
are separated by a month. 

The data used in this work is a subsample of the SVISS, the
simulations have been carried out in the deepest search band
(the $I$ band) and only on one of the four CCDs. The $R$ band
has only been used to determine the spurious detection ratio (see
section~\ref{sec:spur}). Table~\ref{table:data} contains an overview
of the $I$ band data used in the tests. It should be noted that the
total number of frames in the reference epoch is $\sim 80$, we have
only used the best frames to construct the reference frame in order
to optimise the resulting seeing. The large number of frames
available for making a reference image is not by design, most of the
frames were obtained in a previous failed observation period (the VIMOS
instrument was taken off-line due to technical problems in fall 2003).
In this paper we do not discuss situations where the reference epoch
has worse seeing than the search epochs. In the full SVISS
data set we have some epochs with better seeing than the reference
epoch. However, we have not used those epochs in the tests reported in
this paper. More details on the data and data reduction, along with
source catalogues for the fields will be given in a forthcoming paper
(Mencia-Trinchant et al. in prep).

\begin{figure*}
\centering 
\includegraphics[width=0.95\textwidth]{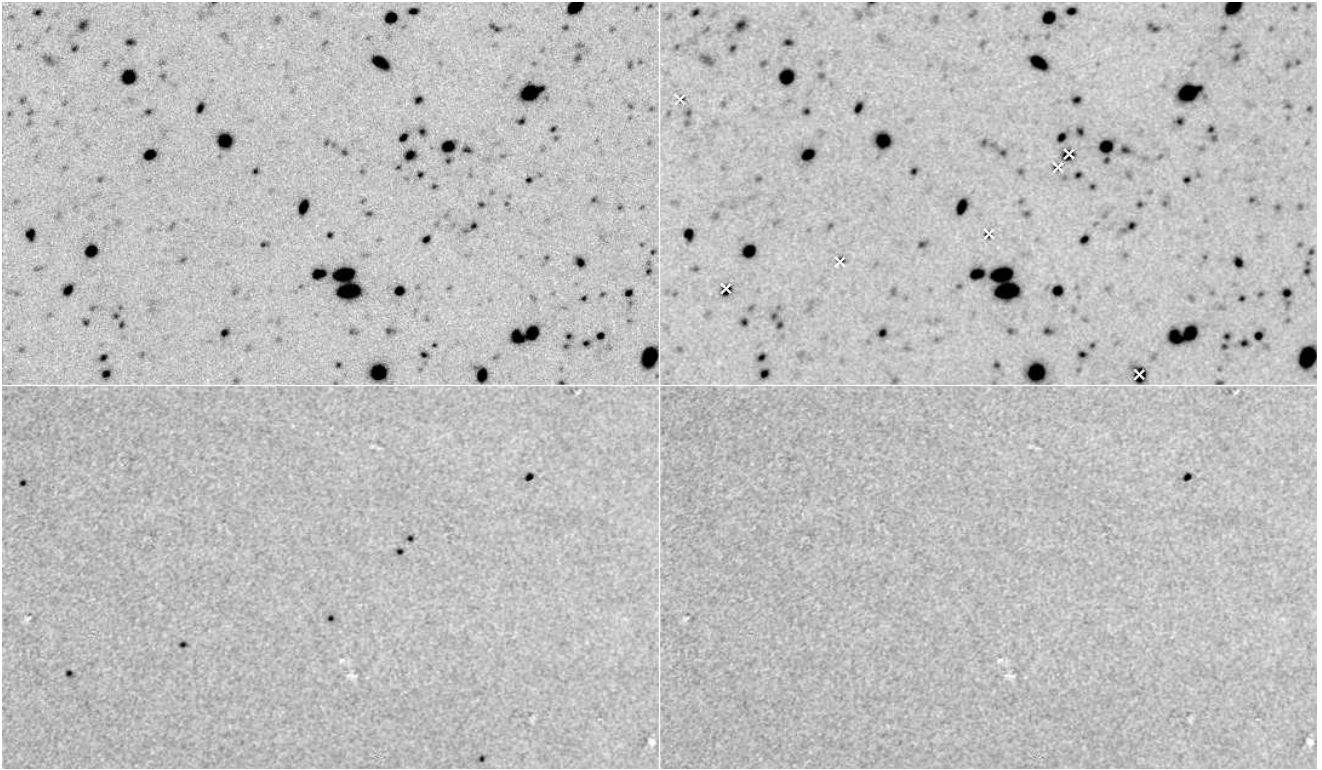} 
\caption{An example image from the SVISS data set containing a number
	of simulated supernovae. The field of view in the image is
	1.9 \arcmin $\times$ 1.1 \arcmin (the total field of view of
	one VIMOS CCD is $\sim$ 7 \arcmin $\times$ 8 \arcmin). The
	top left frame is the reference image, the top right frame
	is the epoch 2 image with supernovae added (marked with with
	white crosses). The bottom left frame is the subtracted image
	(from run A0), the remaining sources are the added SNe. Some
	residuals are visible in the subtracted frame where no SNe
	have been added to the image (bottom right frame).}
\label{figure:simul} 
\end{figure*} 

\begin{table} 
\caption{Data sample overview}	     
\label{table:data} 
\centering 
\begin{tabular}{l c c c l} 
\hline\hline 
Data set/Epoch & Exp. time (s) & $m_{lim, rms}\dagger$ & 
$m_{lim, sim}\ddagger$ & Seeing (\arcsec)\\ 
\hline 
   SVISS/Ref & 11520 & 26.36 & 26.1 & 0.61 \\ 
   SVISS/A & 11520 & 26.64 & 26.3 & 0.68 \\ 
   SVISS/B & 9600 & 26.48 & 26.1 & 0.72 \\ 
   SVISS/C & 14400 & 26.62 & 26.3 & 0.64 \\ 
   SVISS/D & 7680 & 26.61 & 26.2 & 0.65 \\
\hline					 
\end{tabular} 
\\ 
\raggedright $\dagger$ The limiting magnitudes are
3$\sigma$ limits and have been measured for an aperture of diameter
4$\times$ the FWHM.\\
$\ddagger$ The limiting magnitudes are 3$\sigma$
limits and have been estimated from the photometric scatter
of simulated sources in the individual unsubtracted epoch images.
\end{table}

The data were reduced using a data reduction pipeline for SVISS
developed by our team. The data were bias subtracted, flatfielded
and scaled to correct for extinction. The images were also normalised
to counts per second. A bad pixel mask was produced for each of the
images, containing vignetted ($<5$\% in most cases), saturated and
cosmic ray flagged pixels.

The VIMOS $I$-band suffers from rather severe effects of fringing,
and care has to be taken to successfully remove these. We
constructed a fringe map for each observation night by median
combining the non-aligned science frames and rejecting bright pixels
by a standard sigma-rejection routine. The fringe map was then
subtracted from each of the flatfielded science frames. All of the
science frames were also manually inspected, some frames were removed
because of vignetting and/or poor seeing (see Table~\ref{table:data}
for the total number of frames used in each epoch).

The good quality frames for each epoch were registered to a common
pixel coordinate system using a simple shift and rotate transform
(yielding a typical rms of $\sim0.1$ pxls/0.02\arcsec). Finally, the
registered frames were median combined. The bad pixel maps for each
individual image were also registered and summed, yielding an exposure
time map for each epoch. Some pixels ($< 0.2 \%$ in the reference
epoch) had an exposure time of 0, i.e. none of the individual frames
contained useful data for that pixel. These pixels were flagged in
the combined science frames.

Each of the combined epoch images were photometrically calibrated 
using secondary standard stars in the field. The calibrated photometry
for these stars were obtained from stacked one-night images from
nights where standard star observations were available.

For the reference image, a host galaxy catalogue was constructed
using SE. A total of 2660 galaxies were detected, rejecting
all sources with photometric errors larger than 0.1 mag
and sources close to masked regions. For each of the galaxies a flux
profile was calculated for four radii
containing respectively 20\%, 40\%, 60\% and 80\% (hereafter $r_{20}$,
$r_{40}$, $r_{60}$ and $r_{80}$) of the total flux as measured by
SE's AUTO photometry.

The search epochs were aligned to the reference epoch following
the procedure outlined in section~\ref{sec:regmeth}. To obtain
the necessary accuracy for the registration we needed to use a
transformation that allowed separate shifts, rotations and scales for
the x and y coordinates. The reference sources for registration were
bright, point-like objects in the field (approximately 100 sources
were used).  The resulting rms of the geometrical transformations were
smaller than 0.1 pixels for all the epochs.

The subtraction pipeline has a number of input parameters, which need
to be adapted to the data set. Table~\ref{table:params} contain the
list of parameters and the values for the SVISS data set. Since the
VIMOS images are quite large we found it preferable to divide
the images into subregions before the actual subtraction, which
makes it possible to use a low degree of spatial variation in the
kernel. The total number of stamps given in Table ~\ref{table:params}
is for the entire image. If the number of stamps was too small
the subtraction quality degraded considerably. Since the number
of stamps need to be tuned to the particular data set, we do not
report any tests on varying this parameter. The parameters related
to the size of the stamp regions have not been found to affect the
subtraction quality significantly. However, no sub-FWHM sizes have
been considered. The parameters given in Table~\ref{table:params} 
have been kept constant in all the reported tests.

Figure~\ref{figure:snsamp} show two supernova candidates detected
in the $I$ band using the subtraction parameters discussed above. The
bright SN candidate (upper panel in the figure) has an approximate
$I$ band magnitude of 22.7 at peak and is situated inside a bright
, tentatively elliptical, galaxy. The faint candidate (lower panel)
has an approximate $I$ band magnitude of 24.8 at detection, and the
explosion date for this candidate is not very well constrained. The
host galaxy for this SN is fainter and more representative of
average galaxies in our images. These detections demonstrate that the
use of our subtraction pipeline enables us to find faint supernovae
against the bright background emission from the host galaxies.

\begin{table*} 
\caption{Properties and results of the test runs using the SVISS data set.} 
\label{table:vimruns}      
\centering                          
\begin{tabular}{l c c c c c c c c}        
\hline\hline                 
Run & Stamp selection& Host brightness& Kernel widths&
$\sigma_{r5}$/$\sigma_{r1}$& $\chi^2_{norm}$& $df$ & $r_{sp}$ \\    
\hline 
A0  &  High SNR  & $I\lesssim 25$ & (1.0,2.0,4.0)  &0.026/0.034 & 1.00 & 0.13 & 0.825 \\ 
AS1 &  Auto  & $I\lesssim 25$ & (1.0,2.0,4.0)      &0.024/0.031 & 2.97& 0.22 &  0.848 \\ 
AS2 &  Point & $I\lesssim 25$ & (1.0,2.0,4.0)      &0.024/0.034 & 2.87 & 0.12 & 0.891 \\ 
AH1 &  High SNR$\dagger$ & $I\lesssim 22$ & (1.0,2.0,4.0)  &0.027/0.077 & 7.34 & 0.47 & 0.825 \\ 
AH2 &  High SNR& $22\lesssim I \lesssim 24$ & (1.0,2.0,4.0)  &0.027/0.038 & 1.76 & 0.27 & 0.825 \\ 
AW1 &  High SNR& $I\lesssim 25$& (0.2,0.4,0.8)  &0.027/0.033 & 1.39 & 0.13 & 0.831 \\ 
AW2 &  High SNR& $I\lesssim 25$& (0.4,0.8,1.6)  &0.026/0.033 & 0.924 & 0.13 & 0.821 \\ 
AW3 &  High SNR& $I\lesssim 25$& (0.6,1.2,2.4)  &0.026/0.032 & 0.855 & 0.10 & 0.816 \\ 
AW4 &  High SNR& $I\lesssim 25$& (2.0,4.0,8.0)  &0.024/0.034 & 1.88 & 0.11 & 0.834 \\ 
AW5 &  High SNR& $I\lesssim 25$& (4.0,8.0,16.0) &0.027/0.035 & 1.02 & 0.14 & 0.826 \\ 
\hline
B0  & High SNR & $I\lesssim 25$ & (1.0,2.0,4.0)& 0.023/0.035 & 1.00 & -0.19 & 0.667 \\
BS1 &  Auto  & $I\lesssim 25$ & (1.0,2.0,4.0)  & 0.025/0.034  & 1.25 & -0.32 & 0.726 \\ 
BS2 &  Point & $I\lesssim 25$ & (1.0,2.0,4.0)  & 0.026/0.036  & 1.34 & -0.29 & 0.808 \\ 
BH1 & High SNR$\dagger$ & $I\lesssim 22$ & (1.0,2.0,4.0)  & 0.024/0.084 & 2.91 & -0.088 & 0.667\\
BH2 & High SNR& $22\lesssim I \lesssim 24$ & (1.0,2.0,4.0) & 0.024/0.041 & 1.06 & -0.28 & 0.667\\
BW1 &  High SNR& $I\lesssim 25$& (0.2,0.4,0.8)  &0.024/0.034 & 0.638 & -0.030 & 0.648 \\ 
BW2 &  High SNR& $I\lesssim 25$& (0.4,0.8,1.6)  &0.023/0.033 & 0.823 & -0.13 & 0.659 \\ 
BW3 &  High SNR& $I\lesssim 25$& (0.6,1.2,2.4)  &0.023/0.033 & 1.04 & -0.20 & 0.672 \\ 
BW4 &  High SNR& $I\lesssim 25$& (2.0,4.0,8.0)  &0.024/0.035 & 1.06 & -0.22 & 0.677 \\ 
BW5 &  High SNR& $I\lesssim 25$& (4.0,8.0,16.0) &0.026/0.046 & 0.672 & -0.14 & 0.661 \\ 
\hline
C0  & High SNR & $I\lesssim 25$ & (1.0,2.0,4.0)   & 0.026/0.034  & 1.00 & -0.21 & 0.686\\
CS1 &  Auto  & $I\lesssim 25$ & (1.0,2.0,4.0)  & 0.027/0.035  & 0.83 & -0.18 & 0.674\\
CS2 &  Point & $I\lesssim 25$ & (1.0,2.0,4.0)  & 0.032/0.041  & 1.32 & -0.25 & 0.808\\
CH1 & High SNR$\dagger$ & $I\lesssim 22$ & (1.0,2.0,4.0)  & 0.028/0.070 & 2.92 & -0.69 & 0.686\\
CH2 & High SNR& $22\lesssim I \lesssim 24$ & (1.0,2.0,4.0)  & 0.028/0.039  & 1.40 & -0.51 & 0.686\\
CW1 &  High SNR& $I\lesssim 25$& (0.2,0.4,0.8)  &0.026/0.032 & 0.921 & -0.20 & 0.674 \\ 
CW2 &  High SNR& $I\lesssim 25$& (0.4,0.8,1.6)  &0.026/0.034 & 1.19 & -0.25 & 0.693 \\ 
CW3 &  High SNR& $I\lesssim 25$& (0.6,1.2,2.4)  &0.026/0.035 & 0.949 & -0.20 & 0.685 \\ 
CW4 &  High SNR& $I\lesssim 25$& (2.0,4.0,8.0)  &0.026/0.034 & 1.06 & -0.20 & 0.688 \\ 
CW5 &  High SNR& $I\lesssim 25$& (4.0,8.0,16.0) &0.026/0.036 & 1.18 & -0.21 & 0.701 \\ 
\hline
D0  & High SNR & $I\lesssim 25$ & (1.0,2.0,4.0)   & 0.025/0.033  & 1.00 & -0.0053& 0.734 \\
DS1 &  Auto  & $I\lesssim 25$ & (1.0,2.0,4.0)  & 0.025/0.034  & 1.19 & 0.048 & 0.737 \\
DS2 &  Point & $I\lesssim 25$ & (1.0,2.0,4.0)  & 0.029/0.039  & 1.19 & 0.022 & 0.832 \\
DH1 & High SNR$\dagger$ & $I\lesssim 22$ & (1.0,2.0,4.0)  & 0.025/0.069  & 4.00 & 0.21 & 0.734\\
DH2 & High SNR& $22\lesssim I \lesssim 24$ & (1.0,2.0,4.0)  & 0.027/0.038  & 1.10 & 0.099 & 0.734\\
DW1 &  High SNR& $I\lesssim 25$& (0.2,0.4,0.8)  &0.072/0.061 & 2.15 & 0.10 & 0.821 \\ 
DW2 &  High SNR& $I\lesssim 25$& (0.4,0.8,1.6)  &0.030/0.036 & 1.06 & 0.033 & 0.738 \\ 
DW3 &  High SNR& $I\lesssim 25$& (0.6,1.2,2.4)  &0.028/0.036 & 1.18 & 0.071 & 0.738 \\ 
DW4 &  High SNR& $I\lesssim 25$& (2.0,4.0,8.0)  &0.029/0.040 & 1.27 & 0.079 & 0.741 \\ 
DW5 &  High SNR& $I\lesssim 25$& (4.0,8.0,16.0) &0.028/0.040 & 1.66 & 0.044 & 0.749 \\ 
\hline                                   
\end{tabular}
\\
\flushleft
$\sigma_{r5}$ and $\sigma_{r1}$ is the pixel to pixel noise in the
isolated and inside galaxies positions, respectively.
$\chi^2_{norm}$ is the normalised $\chi^2$ statistic for the run, 
$df$ is the flux difference statistic and
$r_{sp}$ is the spurious detection ratio (see section \ref{sec:pixstat} for
details).\\ 
$\dagger$ Bright sources with simulated SNe have been manually removed from the
stamp list.\\
\end{table*}

\section{Simulations}
\label{sec:sim}
\subsection{The supernova point spread function model}
The construction of the artificial PSF
model and the addition of SNe to image frames has been done
with the IRAF package \texttt{daophot} (see also section
\ref{sec:phot}).

For the SVISS data set approximately 40 bright star-like objects
were selected by visual inspection. About half of these objects
were rejected in the fitting procedure. The total number of point
sources used for constructing the PSF model for each image was
thus approximately 20.  The best fitting profile was found to be an
elliptical Moffat profile with $\beta= 2.5$. The
FWHM of the models for different epochs/data sets can be found in 
Table~\ref{table:data}. 

Supernovae of different brightness were simulated in the search images.
In all test runs the SNe were added using the task \texttt{addstar}.
The input model for this task was the best-fitting PSF model and
the stars were added in the search image, i.e. the image with worse
seeing.  A total of 17 magnitude bins were used, the brightest bin
roughly corresponds to the 100$\sigma$ limiting magnitude and the
faintest bin somewhat fainter than the unsubtracted image limiting
magnitude (see Table~\ref{table:data}). The magnitude bins used are:
22.0, 22.5, 23.0, 23.5, 24.0, 24.25, 24.5, 24.75, 25.0, 25.25, 25.5,
25.75, 26.0, 26.25, 26.5, 26.75 and 27.0.

\subsection{Placing the artificial SNe}
\label{sec:simplace}
When adding artificial supernovae to the images care must be taken
that the subtraction process itself is not affected. This puts some
constraints on how many SNe can be placed in each image, and also
where they are placed. Since ISIS relies on the use of stamps, it is
important that there is a sufficient number of sources left without
artificial supernovae. The image subtraction process should, however,
be able to deal with the presence of a supernova inside a previously
selected stamp (and reject it from the stamp list). In a real survey
it is very unlikely that more than one supernova will be found in
the same galaxy during the same survey period. The number of SNe per
galaxy has therefore been limited to one.

With these considerations in mind we are selecting host galaxies at
random from a galaxy list, which, depending on the specific test
run, can contain different samples of the full galaxy catalogue.
The positions of the artificial sources inside the host galaxies are
selected based on the flux profile for the specific host galaxy (in
the general case, some test runs use a slightly different setup). Four
position bins were used, where the in- and outside radii of the
bins are given by: 0 to $r_{20}$; $r_{20}$ to $r_{40}$; $r_{40}$ to
$r_{60}$; $r_{60}$ to $r_{80}$. The angle with respect to the major
axis of the galaxy and the radius within each bin were randomly
selected. A fifth "position bin'' was also considered, namely
isolated positions, i.e. positions in the image without any detected
galaxy within a given isolation radius. These positions were found by
masking the detected sources in the reference image, using a $5\sigma$
cutting threshold, dividing the resulting image into rectangular boxes
with half-sides corresponding to the isolation radius. For the SVISS
data an isolation radius of 20 pixels $\sim$ 4\arcsec was used. The
wanted number of supernova positions were then selected at random
from this list of boxes and sources placed in the middle of the boxes
(one source per box).

In Fig.~\ref{figure:simul} an example of an image with
artificial SNe is shown, together with the subtracted frame. The lower
right panel of the figure shows the subtracted frame without any
simulated sources. Typically, each subtracted frame will 
contain hundreds of residuals similar to the ones in this figure.

\subsection{Tests using the SVISS data set}
\label{sec:tsetup}
We have carried out a number of test runs with simulated supernovae added to
the real images to study the effects on the subtraction quality and SN 
detection. A summary of the properties of the different runs is
presented in Table \ref{table:vimruns} for the SVISS data set.
\begin{figure*}
\centering
\includegraphics[width=16cm]{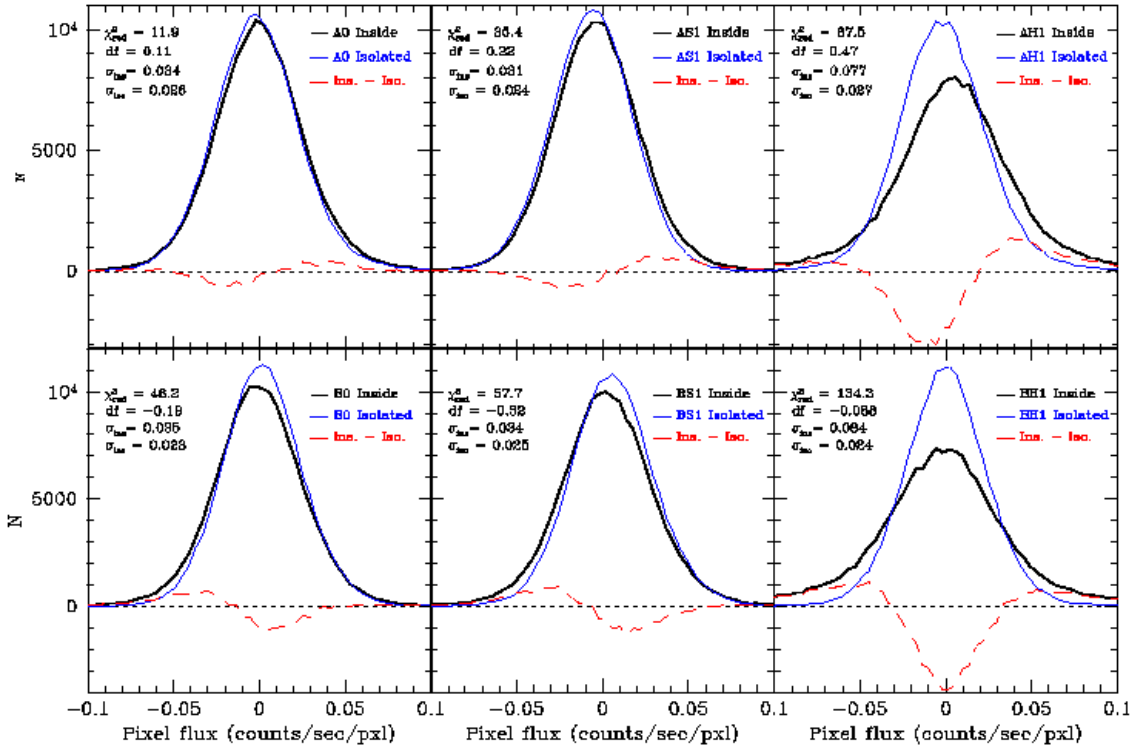}
\caption{Pixel noise distributions (PNDs) for two of the search epochs
(upper panel shows search epoch A, lower epoch B). Leftmost panels
show the reference test runs (using high-SNR stamps and completely
random host galaxies), middle panels show the S1 runs (using ISIS
automatic stamp selection) and rightmost panels show H1 runs (bright
host galaxies). All distributions, except the bright host galaxy
test, contain the same total number of pixels (for details see Table
\ref{table:params} and section \ref{sec:pixstat}).  The dashed (red)
line show the difference of the inside and outside pixel distributions
for each case.  (This figure is available in colour in the electronic
version of the article)}
\label{figure:pixsp}
\end{figure*}
The tests have been done using four different epochs (the A, B, C and D
designations, see Table \ref{table:data} for basic properties of the
different epochs). The reference runs for each epoch, A0-D0, are tests
using an identical set of parameters. For each epoch we have then 
changed one parameter at a time to study its effect on the subtraction
quality. Some parameters have been studied in many epochs, 
while some have only been studied in the epoch 2 data. The same
reference epoch (epoch 0) has been used for all the runs.

The stamp selection has been done in three different ways (e.g. A0,
AS1 and AS2).  In the ``High SNR'' option only the brightest
sources in the field have been selected as
stamps and in the ``Auto'' option the ISIS internal automatic
stamp selection has been used. For the ``Point'' option, the most
point-like sources have been selected as stamps. The total number of
stamps in each of the three settings was $\sim 450$ over the entire frame
($50$ per subregion).

The tests on host galaxy brightness (i.e. 0, H1-H2) were done by
selecting hosts from a magnitude limited galaxy list. In the H1-H2
runs the number of simulated SNe was a bit lower than in the reference runs,
100 instead of 266. For the ``$I\lesssim22$'' option the 100 brightest
sources were selected as hosts, the mean brightness of these galaxies
was $I = 20.0$. For the ``$22\lesssim I \lesssim
24$'' option 100 sources were selected at random from a magnitude
limited catalogue, the mean brightness of these sources was $I = 23.2$. 
The mean brightness of galaxies in the full host galaxy catalogue was $I
= 23.7$. In the case of the brightest host galaxies, some
stamps had to be removed manually as they were not rejected
by ISIS despite the addition of a bright SN to the stamp region.
Having SN host galaxies as stamps affected the subtraction quality
strongly for the brightest supernova bins.

In the W1-W5 runs, the width of the Gaussian basis functions that
constitute the fitting kernel have been changed. The default ISIS
values for the widths (expressed as the Gaussian $\sigma$ parameter)  
are (1.0,2.0,4.0). In the
test runs we are using both narrower and broader widths. The optimal
kernel should contain a basis function with width that when convolved
with the reference image will produce an image with the same seeing
as the search image. For the SVISS images (see Table~\ref{table:data})
the optimal kernel should thus contain a basis function with $\sigma
\sim 0.4-0.8$.

\begin{figure}
\centering
\includegraphics[width=8cm]{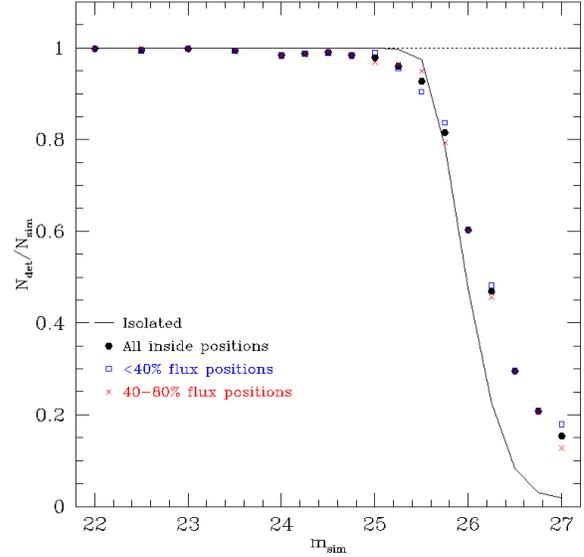}
\caption{Detection efficiency for isolated positions and different
galactocentric distances inside the host galaxies. The plot was
constructed from the A0 run.(This figure is available in colour in
the electronic version of the article)}
\label{figure:posdet}
\end{figure}
\begin{figure*}
\centering
\includegraphics[width=16cm]{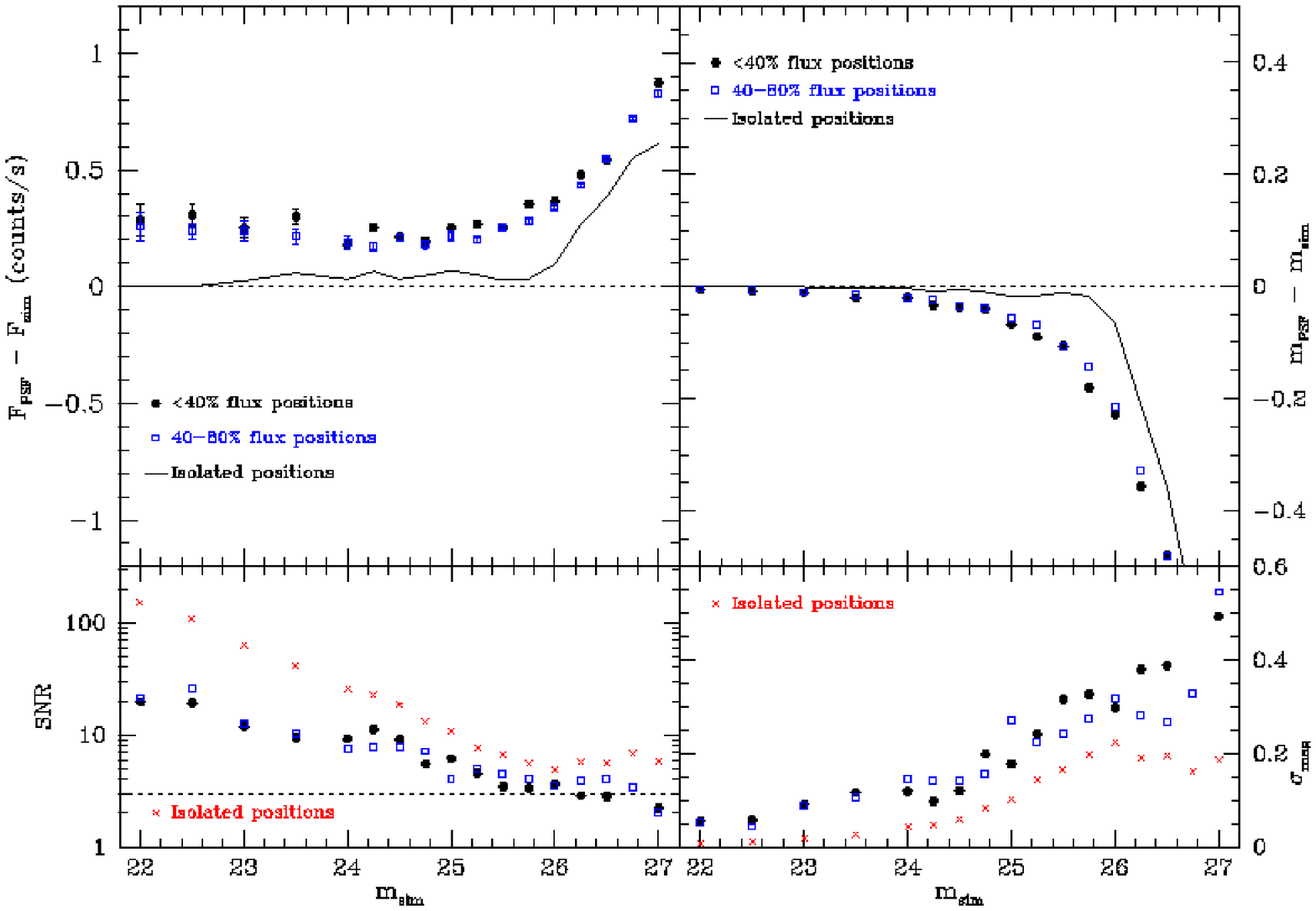}

\caption{Photometric accuracy for SNe at different galactocentric 
distances as a function of simulation magnitude ($m_{\mathrm{sim}}$). 
The plot was constructed from the A0 run. The
	upper left panel shows the median pixel flux difference (in
	counts/sec) between the
	input simulation flux ($F_{\mathrm{sim}}$) and the measured
	PSF flux ($F_{\mathrm{PSF}}$), and the upper right panel
	shows the median magnitude difference. Error bars in the
	flux difference plot show the standard error of mean of
	the difference. Note that the detections become
	incomplete at magnitudes $\gtrsim 26$, this causes the
	apparent failure of photometry at those faint magnitudes.
	The lower left panel shows the signal to noise versus the
	simulation magnitude and the lower right panel the
	measured scatter in each simulation bin. The signal to noise ratios
	have been estimated from the measured scatter ($SNR \approx
	1.0857 \times \frac{1}{\sigma_{\mathrm{mag}}}$). The dashed
	horizontal line marks the $SNR=3$ limit. (This figure
	is available in colour in the electronic version of the
    article)}

\label{figure:posphot}
\end{figure*}
\subsection{Test diagnostics}
\subsubsection{Pixel statistics in subtracted images}
\label{sec:pixstat}
In the AL98 and A00 papers pixel statistics of the subtracted images
are used as a measure of the subtraction quality. The theoretical
pixel noise distribution (hereafter PND) is compared to the measured
distribution in the subtracted frames. The images used in these
subtractions are simulated images with a purely Gaussian PND. The
theoretical distribution is thus quite simple to calculate, even though
convolution of the reference image makes it somewhat more complicated.
In our tests we have used real data images with many sources,
both point sources and extended objects, the PNDs for these images
are harder to model. We have opted to take a more observationally
orientated approach. Instead of comparing the PND in the subtracted
image with a theoretical distribution, we compare the PND of isolated
positions in the subtracted image to that of positions inside galaxies
(see Fig.~\ref{figure:pixsp} for an example of this). The PND for
isolated positions is thus considered to be the optimal one, the
closer the PND of the inside galaxy positions is to the isolated case,
the more optimal the subtraction.

The pixel lists used in the statistics contain all pixels inside a
$3\arcsec$ radius from all of the simulated SN positions, i.e. the
pixel lists are obtained from a list of apertures, measured in an ``empty''
(without simulated SNe) subtracted frame. The positions
for the isolated case are taken from the fifth (isolated) position bin
and the inside galaxies positions from the $r1$ position bin. The same
number of apertures have been used for both pixel lists, so that
the total number of pixels is the same within interpolation errors. 

For each of the test runs we have calculated three different statistics
based on comparing the PNDs: (i) the pixel to pixel noise inside
the apertures of the inside positions and the isolated positions,
$\sigma_{\mathrm{r1}}$ and $\sigma_{\mathrm{r5}}$; (ii) the reduced
$\chi^2$ statistic $\chi^2_{red}$ computed from comparing the PNDs;
(iii) the flux offset statistic, $df$.

The flux offset statistic is obtained by numerical integration of the 
PND difference over the counts per pixel and dividing
by the number of apertures used. The flux offset is defined as:
\begin{eqnarray}
\label{equation:df}
df = \frac{1}{n_{ap}} \!\! \left( \int_{ll}^{0} \!\! - (P_{r1}(x) - 
P_{iso}(x))\,dx \nonumber \right. \\
\left. + \int_0^{ul} \!\! (P_{r1}(x) - P_{iso}(x))\,dx \right),
\end{eqnarray}
where $n_{ap}$ is the number of apertures, $ll$ and $ul$ are the upper
and lower limit of the counts per pixel, respectively, $x$ counts per
pixel and $P_{r1/iso}$ the number of pixels in the interval $dx$ for a
given value of $x$. The $r1$ and $iso$ designations refer to the two position
bins considered (see above). This statistic is in principle the mean
counts per aperture with the background, defined as the counts inside the
isolated position apertures, removed. The numerical integration of
the PND difference enables us to control the lower and higher bounds
as well as making sure that the correct background, i.e. the counts
within isolated position apertures, is removed.

\subsubsection{Detection efficiency}
For each of the magnitude bins a number of detected sources is
obtained from the SE runs. A detected source is defined as having a
signal-to-noise ratio (SNR) of more than one, ``sources'' detected by SE can
have very low significance and include many spurious detections. The
input simulation coordinate list is then used as an ASSOC list in SE
(i.e. matched in image coordinates) to make sure that only simulated
objects are counted in the detection. In this way we can study
the detection efficiency for a certain set of parameters in real
data without including spurious detections from residual sources.
This estimate of the survey depth is useful when other methods can
be used to dispose spurious detections, e.g. by using multiple bands
or epochs. For single epoch measurements greater care has to be taken
not to include spurious objects. 

The detection efficiency is then calculated for each magnitude bin as
the number of detected sources divided by the total number of simulated
objects in the magnitude bin. However, these will not be
the true detection efficiencies since the SNR estimated by SE is not
accurate due to the correlated pixel to pixel noise in the subtracted
images. Such detection efficiencies would thus be unreliable to use
in the SN rate calculations, and in this paper we use them only for
comparison between the different subtraction parameters and epochs. Due
to the way we handle detections in the images with simulated SNe (only
accepting detections at positions where SNe were placed), the signal to
noise cut used does not affect the results discussed in this paper. For
the SN rate calculations we will use detection efficiencies based on
signal to noise cuts found from doing PSF photometry of simulated SNe
(using the photometric scatter method described below).

\subsubsection{Photometric accuracy}
The photometry of the supernova candidates is done in the subtracted
images using PSF photometry (see section \ref{sec:phot}). The accuracy
of this method has been assessed by subtracting the input simulation
magnitudes from the measured magnitudes for each of the simulated
sources and then calculating the mean, median and standard deviation
of the magnitude differences. The magnitudes have also been converted
to flux (in counts/sec) using the magnitude zeropoints. The difference in 
flux and magnitude versus input simulation magnitude have then been
plotted for the different cases. It is important to note that each of
the magnitude bins is independent of each other, the host galaxy and
simulated SN position are selected at random for each of the
bins. Any offset in the the accuracy plots thus means that the entire
image/method suffers from a systematic error, rather than effects from
selecting a ``bad'' sample. Tests have also been done using 
aperture photometry with aperture correction, this gives very similar 
results but with a larger scatter.  

The standard deviation of detected SNe in each magnitude bin
is used as an estimate of the photometric scatter. As discussed in
section~\ref{sec:phot} the error estimates from \texttt{daophot}
and \texttt{phot} will not give accurate photometric errors. The
standard deviation in each simulated magnitude bin should provide a
more accurate estimate for the statistical photometric errors. The
SNR for a given magnitude is computed from
the magnitude scatter using the formula $SNR \approx 1.0857 \times
\frac {1}{\sigma_{\mathrm{mag}}}$.

A problem with estimating photometric
accuracy in this way is that the detected sample will be incomplete at
the faintest magnitudes. Only the very brightest sources in a given
magnitude bin (i.e. sources that happen to be located at the position of
a noise peak or residual flux peak) will be detected at the faint limit.
This can be seen in the photometric accuracy plots, e.g.
Fig.~\ref{figure:posphot}, where the flux difference becomes positive
(indicating that the flux is overestimated) in the faintest magnitude
bins. Conclusions should only be drawn based on the complete bins.

\subsubsection{Spurious detections}
\label{sec:spur}
Whenever source detection codes are used to do automatic detections
there is a risk that some of the detected sources are in fact not real,
but rather noise peaks that are mistaken for real objects. The number
of spurious detections is affected by the detection parameters, a very
conservative setting of these parameters can result in few spurious
detections. However, the problem with using conservative detection
parameters is that real, but faint, sources might be missed. In galaxy
surveys it is important that the number of spurious detections is known
and, in some cases, controlled. Different methods can be used to do
this, one is to run the detection again with the same parameters on
the inverted image. If the noise is Gaussian the amount of spurious
detections is expected to be similar in both the images.

For the subtracted images used in this work this method is not
suitable.  Many of the spurious detections will not be related to the
Gaussian noise in the isolated parts of the image, but
rather associated with residuals from galaxies that were not fully
subtracted.  The ``noise'' in these residuals is not necessarily
Gaussian nor symmetric, thus the inverted image is not a good tool
to use.

We instead define a spurious detection in the subtracted images as a
source that is detected in the test search epoch $I$ band images,
but not detected in the $R$ band images for the same epoch. The
spurious detection ratio is thus defined as:
\begin{equation}
r_{sp} = \frac{n_I - n_{I\cap R}}{n_I},
\end{equation}
where $n_I$ is the number of detected sources in the $I$ band image, 
which is the filter used for all of the tests. $n_{I\cap R}$ is the
number of sources detected in both $I$ and $R$. Source detection 
using SE is done in ``empty'' subtracted images for the different 
parameter sets. For this test the detected sources are 
from the entire image, not only from the simulated positions considered 
for the detection efficiency tests. 
\begin{figure}
\centering
\includegraphics[width=8cm]{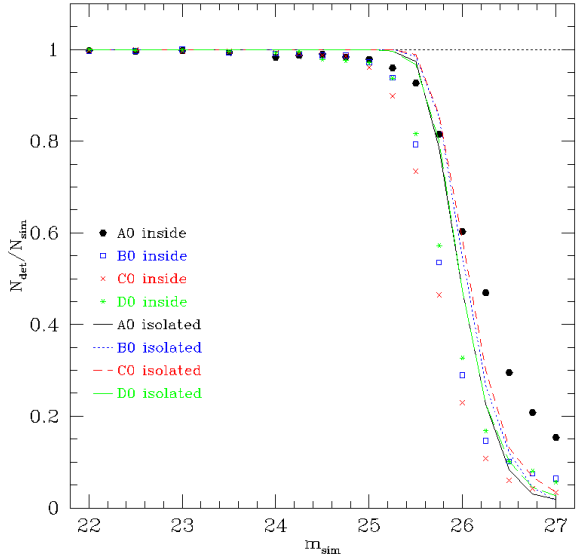}
\caption{Detection efficiencies for isolated and inside galaxies
positions at the four different search epochs.(This
    figure is available in colour in the electronic version of the
    article)}
\label{figure:epdet}
\end{figure}
\begin{figure*}
\centering
\includegraphics[width=16cm]{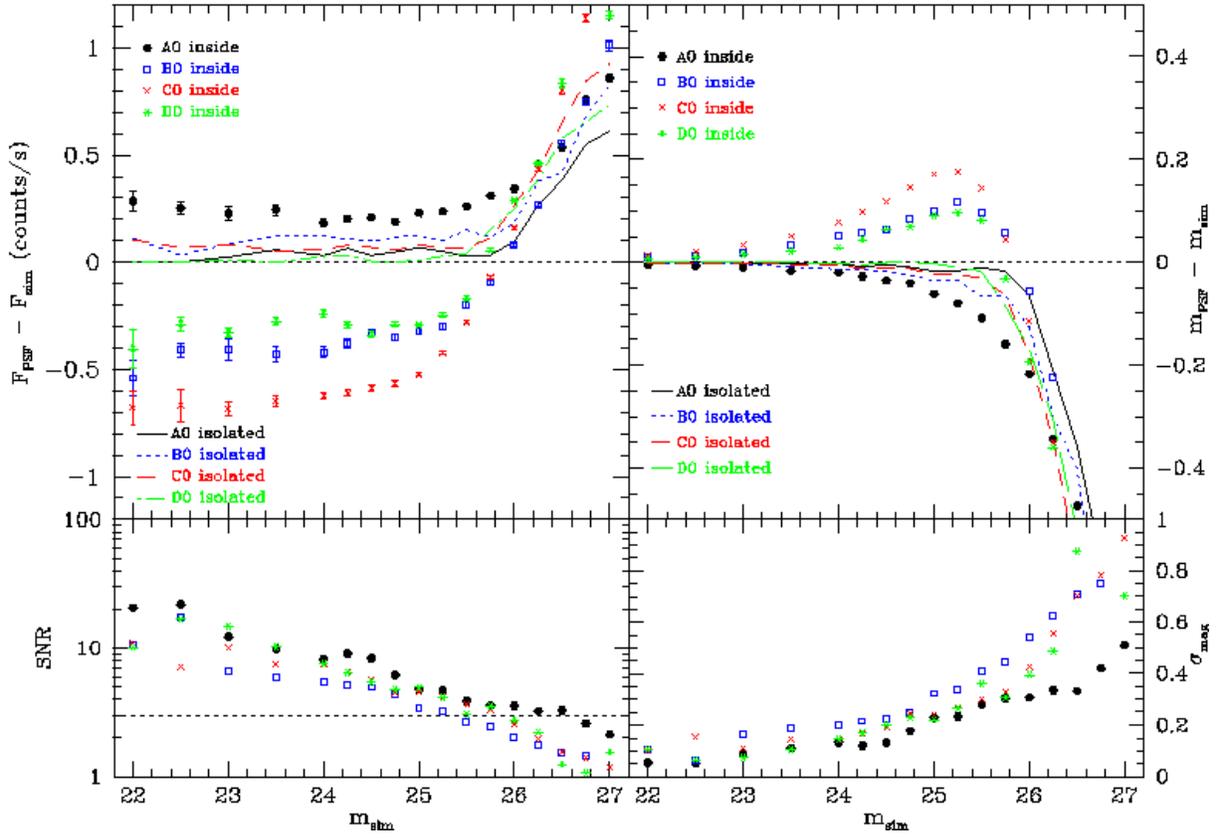}
\caption{Photometric accuracy for the different search epochs. Panels 
    are the same as in Fig.~\ref{figure:posphot}. 
    The legend shows the symbols used for the epochs. (This
	figure is available in colour in the electronic version of the
    article)}
\label{figure:epphot}
\end{figure*}

\section{Results}
\label{sec:results}
Our tests show that the $\sigma_{\mathrm{r1}}$/$\sigma_{\mathrm{r5}}$,
$\chi^2_{red}$ and $df$ statistics can be used as measures on how
successful the difference imaging was in subtracting galaxies from
positions where artificial SNe were added. We have also normalised
the $\chi^2_{red}$ statistic for each test run by the value
found for the reference run (A0, B0, C0 and D0). The normalised
$\chi^2_{red}$ makes it easier to compare the test runs within each
epoch, but can not be used to compare epochs to each other.
Table~\ref{table:vimruns} contains the resulting statistics for
the test runs, and Fig.~\ref{figure:pixsp} shows the pixel noise
distributions for some of them.

Some general remarks can be made on the relationship between the
different statistical measures. By definition $\chi^2_{red}$ will be
a measure of the total difference of the two distributions. There are
two main components to this total difference, one is the difference
in widths of the two distributions (characterised by the standard
deviations, $\sigma_{\mathrm{r1}}$/$\sigma_{\mathrm{r5}}$). The $df$
statistic is mainly sensitive to possible systematic flux offsets,
which is the other main component in the $\chi^2_{red}$, and gives
an indication on whether the offset is positive or negative. In
Table~\ref{table:vimruns} we see that, as expected, a large
$\sigma_{\mathrm{r1}}$ and/or a large $df$ (negative or positive)
results in a large $\chi^2_{red}$. This is also evident from looking
at the pixel distribution plots in Fig.~\ref{figure:pixsp}.

\subsection{Detection and photometry of supernovae with faint host galaxies}
\label{sec:pos}
To test the effects of the supernova galactocentric distance we placed
the simulated sources as described in section~\ref{sec:simplace}. The
resulting detection efficiency and photometric accuracy for the
different position bins are shown in Figs.~\ref{figure:posdet}
and \ref{figure:posphot}. While the position tests have been done
in all the test epochs, these figures show only run A0. The four
position bins inside each galaxy were merged into two, one region
containing 40 \% of the flux and the other one containing 40 to 80 \%
of the flux. We find no difference within the errors between the inside position bins,
neither for detection efficiency nor photometric accuracy. The small
differences visible in the figures is smaller than the scatter for
each magnitude bin. The scatter in the detection efficiency plot
is hard to calculate, but the fact that for values for the two
positions switch order between magnitude bins can be interpreted as
a scatter effect. This result holds true for all of the test runs.

The resulting efficiency and photometric accuracy for the
isolated positions differs significantly from the ones for inside
positions. Assuming the isolated SN positions to be an ideal case and
comparing the photometric accuracy for these positions with the inside
ones, we find that there is a systematic error in the photometry for
the inside positions. Further testing showed that this systematic error
is in fact a flux offset, the difference between simulated and measured
flux is the same in all of the magnitude bins (shown in the left
panels of Fig.~\ref{figure:posphot} and \ref{figure:epphot}). In the
case of run A0 the flux offset is positive, positions inside galaxies
have too much flux remaining after subtraction. This is also visible
in the detection efficiency plot, the efficiency for inside positions
is erroneously boosted by the extra flux, giving an efficiency better
than the isolated case, which is clearly incorrect. The error bars given
in the flux difference panels show the standard error of mean
in each bin. The difference is nonzero (within 3--5 $\sigma$) and is
consistent with a constant flux offset in all of the magnitude bins.
Using the mean flux of the faint host galaxy sample and the
measured flux offset we can calculate the mean fraction of host flux
that remains in the subtracted frame. For the A, B, C and D epochs
the fraction is 1.9 \%, 3.0 \%, 2.5 \% and 5.3 \%, respectively.

The subtraction has succeeded to varying degree in the four different
search epochs, the detection efficiency and photometric accuracy for
the reference runs of each epoch is shown in Figs.~\ref{figure:epdet} and
\ref{figure:epphot}, as well as in Table~\ref{table:vimruns}. The
$\chi^2_{red}$ statistic for the four reference runs are in fact
very different, 11.9, 46.2, 32.7 and 17.3, respectively, for A0-D0.
This is not surprising since the data from the different epochs are
of somewhat differing quality (see Table~\ref{table:data}). The flux
offset has been detected in all of the test epochs to varying degree,
in epoch A it is a positive offset and in B, C and D a negative one.
This is clearly visible both in the trend for the $df$ statistic
and in the photometric accuracy plot (Fig.~\ref{figure:epdet}). In
all of the epochs, a flux offset also results in an artificial
boost/suppression of the detection efficiency. 

The lower panels of Fig.~\ref{figure:posphot} and \ref{figure:epphot}
show the $SNR$ and standard deviation of the measured magnitudes
in each bin for the different positions and epochs. Contrary
to the standard error of mean, the scatter is an estimate of the
photometric uncertainty for individual supernovae. As expected, the
scatter is considerably smaller in the isolated positions compared
to the inside positions. The scatter and $SNR$ for the three epochs
are very similar, but the detection efficiency is different. This
shows that the detection efficiency is affected by the systematic
flux offset. The $SNR$ for a given simulated magnitude bin is higher
for the isolated positions than for the inside positions (lower left
panel of Fig.~\ref{figure:posphot}). This is of course as expected
since the presence of a host galaxy will affect the photometry of
the SN. The estimates of $SNR$ and scatter are unreliable for the
faintest magnitude bins since the sample becomes incomplete in these
bins (e.g., see Fig.~\ref{figure:posdet}). A useful estimate of
the depth can be found by extrapolating from the trustworthy data
points where the detections are 100 \% complete. The $3\sigma$
detection limit for the isolated position is approximately
$m_I=26.1$. This is comparable to the $3\sigma$ limiting
magnitude for the individual frames (see Table~\ref{table:data}). The
increase in noise from subtracting the images which should cause the
limiting magnitudes in the subtracted images to be on the order of
0.4 magnitudes lower is not seen. This is due to the effects discussed
in section~\ref{sec:phot}, where the subtraction actually serves to
remove the problematic background, this effect is roughly similar
(by coincidence) to the increase in pixel to pixel noise. The $SNR$
versus magnitude plots for the four different epoch (lower left panel
of Fig.~\ref{figure:epphot}) show that the B epoch is considerably
less deep than the other three epochs. This is in agreement with the
limiting magnitudes given in Table~\ref{table:data}, where epoch B is
indeed the shallowest of the four epochs. The 50 \% completeness limit
roughly corresponds to the $3\sigma$ limiting magnitude obtained from
the $SNR$ plots, but it should be noted that the flux offsets have
a fairly strong effect on the detection efficiencies. The number of
spurious detections, characterised by $r_{sp}$, also reflects this,
the number is significantly higher in the A epoch (which has a positive
flux offset) than in the epochs with negative offsets.

\subsection{Detection and photometry of supernovae with bright host galaxies}
\label{sec:hosts}
\begin{figure}
\centering
\includegraphics[width=8cm]{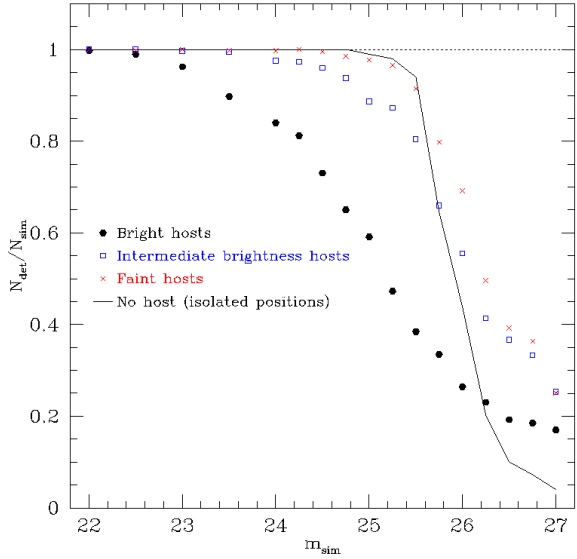}
\caption{Detection efficiency for the runs with different host galaxy
	brightness for a single epoch (A0, AH1-AH2). The mean $I$ band magnitudes
	for the three host galaxy populations are 20.0, 23.2 and 23.7. (This
    figure is available in colour in the electronic version of the
    article)}
\label{figure:hpdet}
\end{figure}

\begin{figure*}
\centering
\includegraphics[width=16cm]{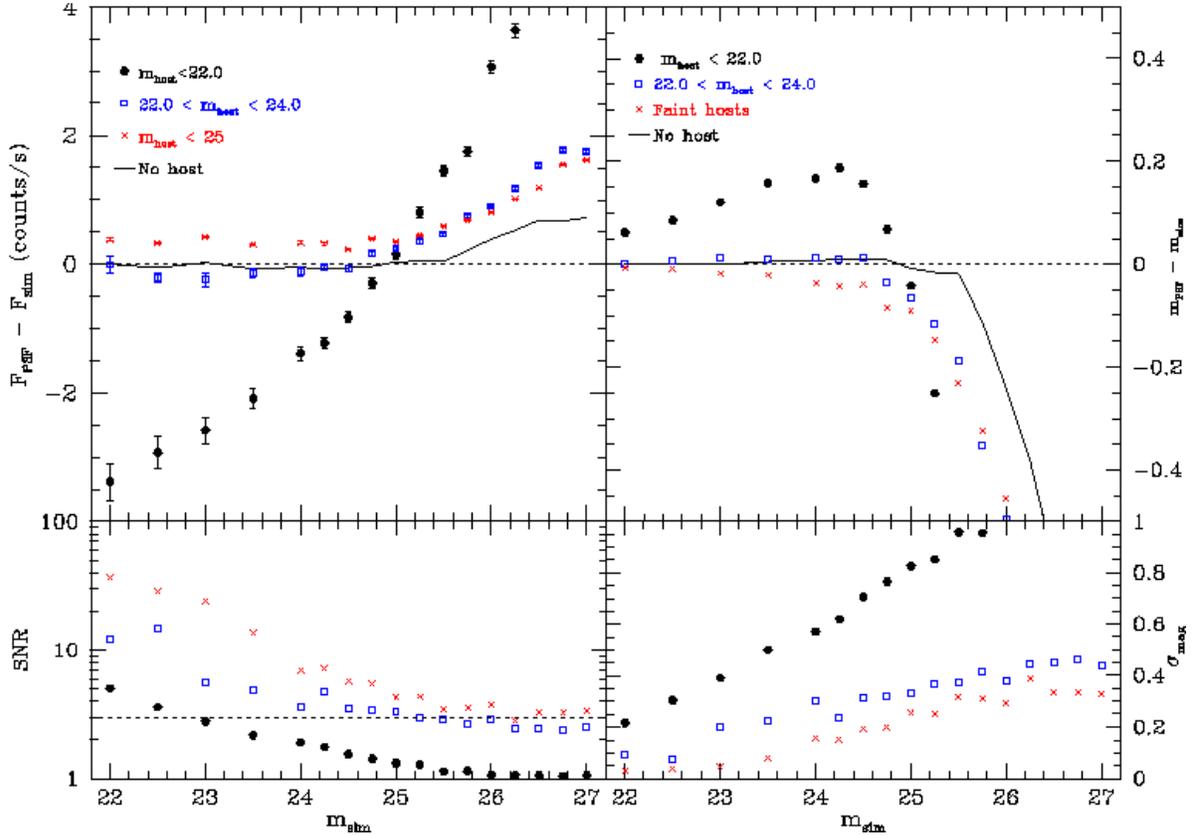}
\caption{Photometric accuracy for runs with different host galaxy 
    brightness for a single epoch (A0, AH1-AH2).Panels 
    are the same as in Fig.~\ref{figure:posphot}. The legends show 
    the symbols used for the different host galaxy tests.
    (This figure is available in colour in the electronic version of the
    article)}
\label{figure:hpphot}
\end{figure*}

Properties of the runs with different host galaxy characteristics can be
found in Table~\ref{table:vimruns} (H1 and H2 runs for the 4 different
search epochs). Figs.~\ref{figure:hpdet} and \ref{figure:hpphot} show
the detection efficiency and photometric accuracy in the reference, H1
and H2 runs for one of the epochs (epoch A). 

The detection efficiency plot clearly shows that the efficiency
becomes worse when the host galaxy is considerably brighter than the
simulated supernovae. This could possibly be explained by the Poisson
noise from the host galaxy eventually becoming dominating when fainter
SNe are detected. Assuming that the noise for a source consists only
of Poisson noise from the host galaxy, i.e. $SNR \sim F_{\mathrm{SN}}
/ \sqrt{F_{\mathrm{host}}}$, we find that the SN magnitude for a 90 \%
detection efficiency corresponds to a signal-to-noise ratio of $\ga
100$ for the bright host galaxy group. Thus it seems unlikely that
the host galaxy Poisson noise affects the detection probability nor
the photometry for the simulated SNe. Also, the drop at the detection
limit is considerably slower for the bright and intermediate brightness
hosts.  These two effects can be explained as subtraction residuals
from the host galaxies being detected as SNe. Visual inspection of the
subtracted frames also show that almost all of the selected bright
host galaxies show significant residuals. The pixel-to-pixel noise
in apertures placed on the inside galaxies position ($\sigma_{r1}$
given in Table ~\ref{table:vimruns}) is also increased substantially
when considering brighter hosts. This is due to the subtraction
residuals in these positions since the Poisson noise from nearly all
but the very brightest galaxies is negligible in our deep and sky
noise dominated images.

The photometric accuracy is also affected by the brightness of the host
galaxies. The constant flux offset discussed in the previous section is
no longer seen for the two populations of brighter host galaxies (see
Fig.~\ref{figure:hpphot}). In the very brightest galaxy population
we see that the flux and magnitude differences are very large.
There seems to be some kind of systematic offset inside the bright
galaxies but it is no longer constant in the different magnitude
bins. For this host sample, the magnitude bins are not independent of
each other, the host galaxies are the same in all the bins, i.e. the
100 brightest galaxies in the field. The flux difference is in this
case not independent of the SN flux. For the very faintest bins the
difference is positive (which also the $df$ statistic for run AH1 in
Table~\ref{table:vimruns} indicates), while for the very brightest
simulation bins the difference is negative. However, it should be
noted that the scatter is very large for the bright hosts. The
SNR plot for different host galaxy brightness (lower left panel of
Fig.~\ref{figure:hpphot}) shows that the 3 $\sigma$ limiting magnitudes
for the bright and intermediate host galaxy samples are a lot worse
than for the faint galaxy sample.

A possible explanation for this effect could be that the photometry
fails in the presence of powerful negative residuals, giving
unpredictable results for the bright simulation bins. The intermediate
brightness hosts are somewhere in between the faint and bright hosts,
for the epoch A case they do not show any offset at all for any of
the magnitude bins. We cannot find any simple relation between the
host brightness and the found photometric difference, the epochs are
affected differently, c.f. AH-DH runs in Table~\ref{table:vimruns}
(Fig.~\ref{figure:hpdet} shows only epoch A).  Since the subtraction
residuals at the bright galaxy locations vary considerably from epoch
to epoch, this is expected if the residuals is the source of
the found photometric errors for bright host galaxies.

It should be noted that the bright host runs were carried out with a slightly
modified stamp list. We found that bright galaxies with simulated
supernovae were not correctly rejected by ISIS in the subtraction. The
subtraction quality was therefore greatly degraded for the brighter magnitude
bins. We manually removed the bright host 
galaxies with simulated SNe from the stamp list. The stamp
rejection is further discussed in section~\ref{sec:sttest}.

\subsection{Effects of convolution kernel properties}
\label{sec:widthresults}
Detailed testing on subtraction parameters in OIS methods have already
been reported by, e.g., A00 and \citet{2007AN....328...16I}. In
these papers the detailed testing is done on fully simulated data
where noise and subtraction residuals can be controlled. We have
done some tests on varying the subtraction parameters in our real data,
i.e., observed images with simulated SNe. In general, the noise
and residuals from the host galaxies dominate over any effects
resulting from subtraction parameter changes. We have therefore
chosen to use the default ISIS 2.2 parameters for all parameters but
the convolution kernel width (see Table~\ref{table:params}). In the
reference runs (the 0 runs) the default kernel widths have been used.

\citet{2007AN....328...16I} find that the widths of the basis functions
of the kernel affect the subtraction result quite strongly. We
have tested this in our real data and the results are shown in
Table~\ref{table:vimruns}. The tests show that the effect of 
changing the widths is small for most of the tested width combinations. 
The $\chi^2_{norm}$ statistic indicate that the widest (W4-W5) kernel
widths are the worst. The best kernel width parameter setting is
different for the four epochs but in many cases kernel width
combinations that are more narrow than the default give a better
result. We have also calculated the detection efficiencies 
for several kernel width combinations. The 0 and W1-W3 runs all have very
similar efficiencies, but for the widest width combinations the
depth goes down considerably. This is consistent with the noise in
the subtracted image ($\sigma_{\mathrm{r1}}$) being $\sim 4$ times
higher when using the widest basis functions (W4-W5) than in the reference
case. In one case (BW5) the $\chi^2_{norm}$ statistic indicates that
this is the best setting for that particular epoch, but the noise in
the subtracted frame is high and the detection efficiency is very poor.
For our data set, where the seeing differences are quite small, it
is likely that the lack of at least one narrow basis function makes
the convolution kernel suboptimal. In principle one would expect the
pixel to pixel noise to go down when using a wider convolution kernel,
but it seems that the higher subtraction residuals in reality
increase the noise substantially in isolated positions and slightly
in the non-isolated positions.

\subsection{Stamp selection}
\label{sec:sttest}
The different stamp selection methods are compared
using the subtraction quality diagnostic statistics in
Table~\ref{table:params}. The tests show that the ``High SNR'' option
works best, both in terms of overall subtraction quality, as estimated
by the $\chi^2_{norm}$ and $df$ diagnostics, and in terms of spurious
detections. The ``Point'' option gives the worst results for all of
the tested epochs. The ``Auto'' option gives differing results, which
is not surprising since the stamp list will be remade for each epoch,
in contrast to the two other methods where roughly the same stamps
are used for all test epochs. In general, this method seems to give
slightly worse results than the ``High SNR'' one.

It is also worth noting that the stamp rejection routine does not
seem to work properly. Galaxies with SNe (bright or faint) are not
always rejected in the stamp selection by ISIS. This is a problem for
the automatic stamp selection method where the subtraction quality
can go down considerably. With the other methods the stamp list can
always be manually edited before the subtraction and the non-suitable
stamps removed from the list.
\begin{figure}
\centering
\includegraphics[width=8cm]{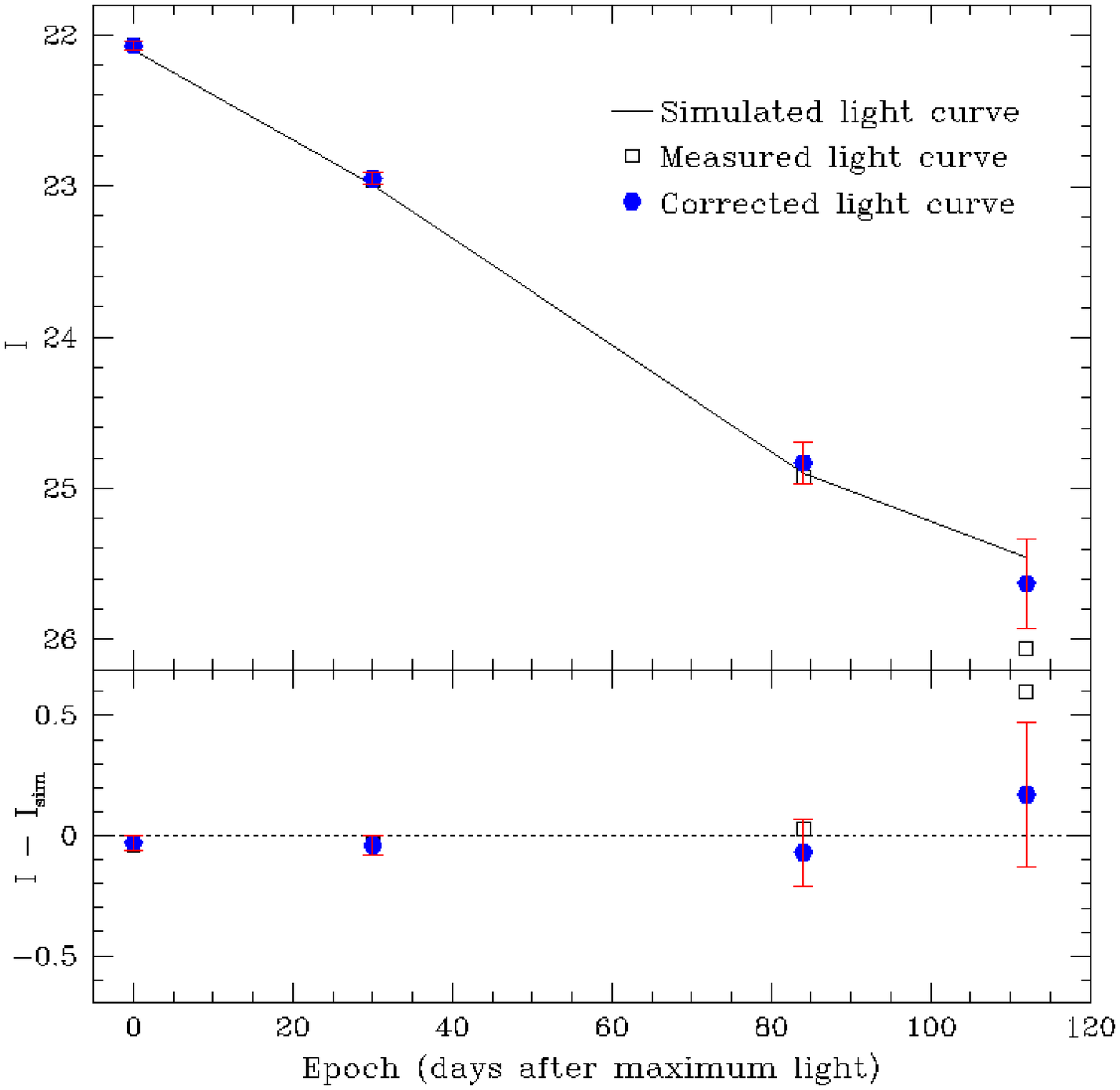}
\caption{``Observed'' light curve of a simulated type Ia supernova at $z = 0.5$. Solid
line is the model light curve in the VIMOS I band from peak magnitude to
$\sim$ 120 days (in the observers frame). Open (black) squares show the measured SN magnitudes in the
subtracted frames and solid (blue) hexagons show the corrected
magnitudes (see text for details). The error bars show the scatter
measured by the detailed simulations reported in
section~\ref{sec:pos}, and are shown as $\pm 1 \sigma$ error bars to
the corrected magnitudes.  For the brightest magnitudes the errors
are smaller than or roughly the size of the corrected points and
thus not visible in the plot. (This figure is available in colour in the electronic 
version of the article)}
\label{figure:lccor}
\end{figure}

\subsection{Estimating photometric errors for SN light curves}
\label{sec:lccor}
Running the detection pipeline on simulated supernovae can provide
useful estimates of possible systematic offsets as well as photometric 
errors. Figure~\ref{figure:lccor} shows the
light curve of a type Ia supernova at redshift 0.5 (light curve model
constructed using the templates provided by P. Nugent\footnote{
http://supernova.lbl.gov/$\sim$nugent/nugent\_templates.html}). The
measurements were done in the subtracted images using PSF fitting
photometry (as discussed in section~\ref{sec:phot}). Using the
simulations reported in section~\ref{sec:pos} we have corrected the light
curve for the systematic flux offset and estimated the uncertainty for each
epoch measurement. Since the pixel to pixel noise in the subtracted
images is correlated the magnitude errors reported by
\texttt{daophot} will not be reliable (they will underestimate the
errors). The uncertainties are instead estimated as the scatter of the
simulated magnitude bins (see figure~\ref{figure:posphot}, interpolating when
necessary). The systematic offset and the scatter will be different for
each epoch, thus simulations need to be carried out in each epoch.
These realistically estimated errors
are on the order of 2 -- 5 times higher than the standard output
from \texttt{daophot}.

The correction for the systematic flux offset is calculated as
a mean quantity for a large number of SNe. For the case shown in
Fig.~\ref{figure:lccor} the corrected magnitudes is a better match
to the simulated light curve than the uncorrected ones, the mean
residual for the uncorrected and corrected curve is 0.175 and 0.0775,
respectively. It is evident that the small flux offset is negligible
for the brighter points on the light curve and it is mainly the
faintest point that is affected by the correction (also the one giving
rise to most of the change in mean residual). It should also be noted
that once the faintest data point has been corrected it is consistent
(i.e. within 1 $\sigma$) with the simulated light curve.

\section{Discussion}
\label{sec:disc}
\subsection{Overall pipeline performance}
We have studied the detection efficiency and photometric
accuracy of our SN detection pipeline using real data from the
SVISS. However, these results can likely be applied to any SN
ground based imaging survey using ISIS subtracted images for detection 
and/or photometry. Previous tests of image subtraction software
have been done mainly on artificial images \citep[AL98, A00,
][]{2002A&A...381.1095G,2007AN....328...16I} but also on stellar
fields (AL98). To our knowledge the work presented in this paper is
the first detailed study of subtraction quality in real supernova
survey data. Detection efficiency studies in supernova survey data
are regularly carried out to be able to find the supernova rates
\citep[e.g.][]{2005A&A...430...83C,2005AJ....130.2272B,2007astro.ph..1043M}, 
but this testing is seldom reported in detail. 

We have found a systematic flux offset affecting the photometry in
subtracted images. In general, with host galaxies chosen at random in
the field, the offset is not dependent on the SN brightness, the offset
is constant for different simulated SN magnitude bins. The offset is
only found for SNe placed inside host galaxies whereas in isolated positions
no offset is found. The offsets change between the different
epochs, in some cases being positive and in others negative.
Pixel statistics for pixels inside apertures centred on the host
galaxies also show that there is a shift of the median pixel value
in the same direction as the flux difference plots indicate and as
expected, a broadening of the distribution. We interpret this as a
small flux through/plateau at the locations of galaxies, reaching
as far out as to the 80\% flux radius. Currently we
do not have an explanation for this effect, it cannot be solved
by changing the convolution kernel sizes or stamp sizes. Further
testing is needed to pinpoint if the systematic offset is caused by
properties of the dataset or if it is in fact an internal effect of
the image subtraction method. It should be noted that this is a small
offset, the systematic error on the magnitudes of SNe is
smaller for brighter sources (for the brightest SNe considered in
this paper, $m_{\mathrm{I}} = 22.0$, the offset is $\sim 0.005$)
and in general significantly smaller than the photometric scatter
for an individual source. However, in a survey of multiple objects,
and particularly fainter objects, this systematic effect will likely
affect the result unless dealt with.

A simple estimate of the photometric uncertainties using the
pixel to pixel noise in the local background underestimates the errors
when doing photometry on SNe in subtracted images. To get an accurate
noise estimate we have instead computed the photometric scatter for
each simulated magnitude bin. This scatter has also been used to
construct signal to noise ratios, using this method the $SNR$ is 
not affected by the flux offset and provides a simple way of rejecting 
less trustworthy observations.

\citet{2007astro.ph..1043M} use simulations to check whether image
registration and subtraction affect their SN photometry.  They find no
systematic offsets to a $SNR$ of 5, but this requires an accurate SN
position obtained from multiple points on the light curve. They find the 
photometric scatter to be
on the order of 20 \% larger than what their noise maps indicate
(but the noise maps already include the effects of re-binning and
convolution). This
seems considerably better than the accuracy we find using our pipeline
and simulations on the SVISS data. The method used by these
authors is quite different from the one
used in this work, therefore it is hard to compare the exact figures. 
We estimate the $SNR$s
differently by using the scatter found from simulations as
our noise. It should also be noted that the SN positions in the images
used by \citet{2007astro.ph..1043M} are chosen at random (private
correspondence G. Miknaitis), thus their results will not be directly
comparable to our inside host galaxies simulations. 
The main difference is that we find that subtraction residuals 
from the host galaxies can have a noticeable effect on the photometric 
accuracy of the survey. This effect does not seem to be present in the
work of \citet{2007astro.ph..1043M}.

The presence of spurious detections is a problem for all supernova
surveys, and in particular for surveys with spectroscopic follow-up
\citep[see the discussion in][]{2007arXiv0705.0493B}. For multi-band,
multi-epoch imaging surveys (like the SVISS) the spurious detections
is less of a problem, but it is still desired to keep the number
down. The bright residuals found in less successful image subtractions
can easily be mistaken for real varying sources when using automatic
source detection. The number of spurious detections in our tests is
large, partly because we use a low detection threshold, but in general
we find that a single epoch, single filter run results in a spurious
detection rate of around 80\% (i.e. 1 out of 5 are real sources)
when the detection threshold is set to find objects brighter than the
5$\sigma$ limiting magnitude. For the SVISS dataset used in this paper
the total number of detected sources for a given epoch is on the order
of hundreds, making the number of spurious detections quite large.
We find that the use of a controlled and bright stamp list makes
the number of spurious detections go down by about 10\%. It should
be noted that residuals at the locations of bright galaxies will be
problematic even in a multi-band survey since the residuals often
appear at the same positions in all filters.

\subsection{Host galaxy brightness and supernova position}
We find that the typical brightness of the host galaxies does
affect the detection efficiencies and photometric accuracy.
This is of course as expected and also confirmed by other SN surveys
\citep[e.g.][]{2006AJ....132.1126N}. However, both detection efficiency
and photometric accuracy becomes gradually worse when considering host
galaxies of increasing brightness. The effect is mainly noticeable for
the very brightest host galaxies (i.e. largest contrast between the
SN and the host). The fact that the photometry is affected by host
brightness can in principle cause the systematic offset found above
to vary with the SN brightness, since the SN and host brightness are
correlated through the host galaxy distance. In projects where the
supernova rate is studied, the effects of host brightness most likely
have an impact on the detection efficiencies at low redshifts, where
the host galaxies can be very bright, as well as in bright galaxies
at high redshifts where the supernovae will be very faint. Whether
this effect will be noticeable depends on the characteristics of
the galaxy population, and simulations of the type reported in this
work can successfully quantify the effect for galaxies of different
brightness. This information can then be incorporated into the overall
detection efficiencies discussed above. The SNe detections
in the SVISS data are likely to be found in faint galaxies and thus
in the general case not affected strongly by the host background flux.

The effect of the SN galactocentric distance in the host galaxy is
difficult to study in the SVISS data set, since most of the galaxies
have small angular sizes, and in many cases are not even resolved. Our
results indicate that the flux offset is slightly worse in the inner
parts (40\% in flux units) of the galaxies than in the outer (40-80\%)
regions. Therefore, we cannot draw any conclusions on how the detection
efficiencies are affected by the galactocentric distances. However,
if the detection efficiency is dependent on radial distance from
the host centre, the radial distribution function of SNe needs to be
known to find a proper correction factor for the total efficiency.

\subsection{Image subtraction method}
Of the different tested parameters we find that the stamp selection
method affects the subtraction quality and detection success the most.
The automatic stamp selection in ISIS mostly works fine, but the lack of
control in selecting stamps can be problematic when dealing with
real SN survey data, due to the presence of saturated objects and
the rejection routine sometimes failing for varying objects. Using a set of
sub-optimal stamps will result in a subtracted frame with worse subtraction
quality.  We note that the optimal selection for stamps is based on SNR, 
selecting preferentially point-like sources as stamps,
i.e. in general fainter sources than in the SNR based selection,
degrades the subtraction quality. Varying the widths of the Gaussian
basis functions of the kernel within a reasonable range does not significantly
alter the detection results, however it should be noted that if all
of the functions are too wide, depending on the data in question the
image subtraction will fail.We also note that for surveys with small
seeing differences, like the SVISS, the default kernel widths used in
ISIS might be to wide. Using widths more in line with the
theoretically optimal convolution kernel width can marginally
improve the subtraction quality.

\subsection{Implications for SN surveys}
The detection efficiency of supernovae in a multi-epoch survey
can be studied by Monte Carlo simulations of SN light curves
as observed by the telescope used for the survey \citep[see
e.g.][]{2004ApJ...613..189D}.  The light curves, together with
the individual detection efficiencies for each epoch can then be
used to find an overall detection efficiency for supernovae of a
certain type and at a given redshift. The flux offset effect found
in this work will affect the detection limits as well as the number
of spurious detections. These can be handled in different ways, but
their presence will always be a problem and minimising the number of
spurious detections is frequently important. A positive flux offset
will result in an artificially boosted detection efficiency and an
increase in the number of spurious detections. A negative offset
will instead cause the detection efficiency to go down, making
the survey to lose depth. This effect can possibly be countered by
controlling the detection threshold used by SE using a weight map,
in principle redefining the threshold for positions considered to be
inside galaxies.

It is very unlikely that systematic flux offsets of the kind found
in this work affect the results found by the large SN Ia projects. In
these projects \citep[e.g.][]{2006ApJ...637..427B,2006A&A...447...31A,
2007astro.ph..1043M}, the photometric accuracy is studied extensively
to make sure that the systematic errors in the cosmological
parameter determination is minimised. As shown above the detection
efficiency will be influenced by the flux offset. Projects studying
the supernova rate and doing detections in images subtracted
using methods similar to ours might suffer from this effect
and detailed simulations should be run to make sure that no
offset is present (or correct for it otherwise). The results from some
recent surveys \citep{2004ApJ...613..189D,2005A&A...430...83C,
2006ApJ...637..427B} are based on supernovae
with observed spectroscopy. The limiting magnitudes for these surveys
are thus set by the magnitude limit of the spectroscopic instrument used
, which is considerably brighter than the corresponding imaging limit, 
thus the offset problem described here is likely not an issue. However, with
present and upcoming surveys based on imaging alone and pushing the
detections to the fainter limits this effect needs to be understood and
studied thoroughly.

Based on the findings reported in this paper we can make some
recommendations for supernova detection and photometry using image
subtraction methods.  Testing should be done on at least part of the
data to determine optimal parameters for the image subtraction. The
stamp selection should be controlled in some way, a good option
is to create a master stamp list from the reference image and use
that in all subtractions but making sure that none of the stamps
contain SNe. If the automatic stamp selection in ISIS is used, care
must be taken that the selected stamps are not variable or contain
saturated pixels. To make sure that a possible flux offset will not
affect the results and to estimate photometric errors, simulations
should be done. If photometry is done in the subtracted frames, the
measurements for an individual SN should be carefully done. This can
be done by simulating SNe of similar magnitude in nearby galaxies of
the same brightness and size as the host galaxy and by measuring the
possible offset. The photometric errors for each point on the light
curve should be estimated by using the magnitude scatter found from
simulations. If no offset correction for the detection efficiencies is
applied we recommend that the detection efficiencies should be capped
by the efficiency for isolated SNe (in the presence of a positive flux
offset), the true efficiency can never be better inside than outside
galaxies.  

\section{Summary} 
\label{sec:summary} 
We have described
the SN detection and photometry pipeline used for the SVISS. We
have also presented detailed testing of supernova detections and
photometry using image subtraction and automatic detection. Our main
results and conclusion are: 
\begin{itemize} 
\item{Image subtraction
using ISIS 2.2 works well for the SVISS, the subtraction quality as
measured by noise levels in the location of galaxies is in general
good. Simulations show that our pipeline is successful in detecting
SNe as well as in doing photometry on them.} 
\item{The detection
efficiency and photometric accuracy for supernovae in a single epoch
is affected by how the stamp selection for ISIS is done.  We have also
found that the host galaxy brightness does affect these quantities,
likely through the presence of powerful subtraction residuals at
the location of bright galaxies.} 
\item{Image subtraction using ISIS
benefits from using a controlled stamp list. We have found that the
best subtraction quality is achieved when the subtraction stamps used
have high signal-to-noise ratios, ideally the brightest non-varying
sources in the field should be used as stamps. The number of stamps
depends on the size and quality of the image, and for our data set
$\sim$60 stamps per ($1k\times1k$) pixels has proven to be sufficient.}
\item{We have discovered a systematic flux offset in subtracted
images. This offset only appears at the locations of galaxies and
is quite small. Photometry of simulated SNe in isolated positions
show no offset. For individual sources the offset is smaller than the
estimated photometric scatter. This effect can cause systematic errors
in the photometry if not corrected for. Detection limits of SNe will
also be affected by non-zero offset, causing problems with increased
numbers of spurious detections for positive offsets and a decrease in
survey depth for a negative offset.  It should be noted that, based
on the tests run so far, we cannot say for sure whether the offset is
something unique to the SVISS data set or if it is due to the image
subtraction method.} 
\item{Simulations of supernovae both
in and outside galaxies are needed to obtain proper estimates of the
photometric errors. The simulations can also be used to correct for
possible systematic offsets.} 
\item{Using automatic source detection
(in this case SE) we find that the detection efficiencies can be
pushed to the limits of the data. The number of spurious detections
in a single epoch subtraction is very large, on the order of 80\%,
but decreases as a function of subtraction quality. To reject spurious
detections other methods must be used (e.g. multiple filters, light
curve fitting).} 
\end{itemize} 
Understanding possible systematic
effects of the subtraction and detection process is very important
when interpreting the results from the SVISS, but will also be
imperative for the upcoming supernova projects based on large imaging
surveys, e.g.  Pan-STARRS\footnote{http://pan-starrs.ifa.hawaii.edu},
LSST\footnote{http://www.lsst.org/lsst\_home.shtml} and SNAP
\citep{2005NewAR..49..346A}.  

\begin{acknowledgements} We would like to
thank J. Sollerman, M. Hayes, Tomas Dahlen for helpful conversations
and comments. We thank P. Price for useful discussions on the use of
the image subtraction code. We would also like to thank the anonymous
referee for many helpful and insightful comments and suggestions. Part
of this work was based on observations done in service mode at the
VLT/VIMOS instrument at Paranal Observatories.

We are grateful for financial support from the Swedish Research 
Council. S.M. acknowledges financial support from the Academy 
of Finland (project:8120503). 
\end{acknowledgements}

\bibliographystyle{aa}
\bibliography{database}
\end{document}